%% file: spt.tex
\documentclass[11pt]{article}
\RequirePackage{algorithm,algorithmic,amssymb,epsf,epic,amsmath}

\input{latexmac}
\denseformat
\usepackage{multirow}
\usepackage{boldline}

\usepackage{pdfsync,fullpage,comment}
\usepackage{authblk}

\ifx\pdftexversion\undefined
\usepackage[colorlinks,linkcolor=black,filecolor=black,citecolor=black,urlco
lor=black,pdfstartview=FitH]{hyperref}
\else
\usepackage[colorlinks,linkcolor=blue,filecolor=blue,citecolor=blue,urlcolor
=blue,pdfstartview=FitH]{hyperref}
\fi

\newcommand{\alert}[1]{\textbf{\color{red}
[[[#1]]]}\marginpar{\textbf{\color{red}**}}\typeout{ALERT:
\the\inputlineno: #1}}

\newcommand{\namedref}[2]{\hyperref[#2]{#1~\ref*{#2}}}

\begin{document}
\def\hpi{\hat{\pi}}
\def\homega{\hat{\omega}}
\def\rt{\mathit{rt}}
\def\hd{\hat{d}}
\def\chS{\hat{\cal S}}
\def\chP{\hat{\cal P}}
\def\chU{\hat{\cal U}}
\def\exp{\mathit{exp}}
\def\Lab{\mathit{Label}}
\def\Id{\mathit{Id}}
\def\Lev{\mathit{Level}}
\def\URt{\mathit{URoots}}
\def\Ball{\mathit{Ball}}
\def\wmax{{w_{max}}}
\def\tO{\tilde{O}}
\def\nin{\not \in}
\def\emset{\emptyset}
\def\setmns{\setminus}
\def\etal{{et al.~}}
\def\Pairs{\mathit{Pairs}}
\def\Paths{\mathit{Paths}}
\def\pred{\mathit{pred}}
\def\Rad{\mathit{Rad}}
\def\succ{\mathit{succ}}
\def\NULL{\mathit{NULL}}
\def\exp{\mathit{exp}}
\def\Ball{\mathit{Ball}}
\def\tPi{\tilde{\Pi}}
\def\deg{\mathit{deg}}
\def\TB{\cB^{{1/3}}}
\def\Branch{\mathit{Branch}}
\def\uzero{u^{(0)}}
\def\uone{u^{(1)}}
\def\uj{u^{(j)}}
\def\vzero{v^{(0)}}
\def\vone{v^{(1)}}
\def\vj{v^{(j)}}
\def\dzero{d^{(0)}}
\def\done{d^{(1)}}
\def\dj{d^{(j)}}
\def\di{d^{(i)}}
\def\hi{h^{(i)}}
\def\diplus{d^{(i+1)}}
\def\hiplus{h^{(i+1)}}
\def\deltaiplus{\delta^{(i+1)}}
\def\dG{d_G}
\def\hsigma{\hat{\sigma}}
\def\third{{1 \over 3}}
\def\stretchexp{{\log_{4/3} 7}}
\def\tLambda{\tilde{\Lambda}}
\def\tomega{\tilde{\omega}}
\def\HKNfactor{ 2^{\tO(\sqrt{\log \max \{n,\Lambda\}})}}
\def\ourfactor{ (1/\eps)^{O(\sqrt{{\log n} \over {\log\log n}})} \cdot 2^{O(\sqrt{\log n \cdot \log\log n})} }

\def\ID{\mathit{Id}}
\def\sqrk{(\sqrt{n}k)}

\def\StagGi{{S'}_G^{(i)}[k]}
\def\StagGiplus{{S'}_G^{(i+1)}[k]}
\def\Stagsqrt{{S'}_G^{(\sqrt{n} \cdot k)}[k]}
\def\Stagtagi{{S'}_{G'}^{(i)}[k]}
\def\Stagtagk{{S'}_{G'}^{(k)}[k]}
\def\Stagtag{S'_{G'}[k]}

\def\Stagkk{{S'}_G^{\sqrk}[k]}
\def\Gtagk{{G'}^{(k)}}
\def\CONGEST{\mathit{CONGEST}}
\def\tOmega{\tilde{\Omega}}

\newcommand{\Patrascu}{P\v{a}tra\c{s}cu{~}}
\newcommand{\Lists}{{\rm Lists}}
\newcommand{\td}{{\tilde{d}}}

\title{Distributed Exact Shortest Paths in Sublinear Time}
\author[1]{Michael Elkin\thanks{This research was supported by the ISF grant No. (724/15).}}

\affil[1]{Department of Computer Science, Ben-Gurion University of the Negev,
Beer-Sheva, Israel. Email: \texttt{elkinm@cs.bgu.ac.il}}

\date{}
\maketitle

\begin{abstract}
The distributed single-source shortest paths problem is one of the most fundamental and central problems in the message-passing distributed computing. Classical Bellman-Ford algorithm solves it in $O(n)$ time, where $n$ is the number of vertices in the input graph $G$. Peleg and Rubinovich \cite{PR99} showed a lower bound of $\tOmega(D + \sqrt{n})$ for this problem, where $D$ is the hop-diameter of $G$. 

Whether or not this problem can be solved  in $o(n)$ time when $D$ is relatively small is a major notorious open question. Despite intensive research \cite{LP13,N14,HKN15,EN16,BKKL16} that yielded near-optimal algorithms for the {\em approximate} variant of this problem, no progress was reported for the original problem. 

In this paper we answer this question in the affirmative. We devise an algorithm that requires $O((n \log n)^{5/6})$ time, for $D = O(\sqrt{n \log n})$, and $O(D^{1/3} \cdot (n \log n)^{2/3})$ time, for larger $D$. This running time is sublinear in $n$ in almost the entire range of parameters, specifically, for $D = o(n/\log^2 n)$. 

We also generalize our result in two directions. One is when edges have bandwidth $b \ge 1$, and the other is the $s$-sources shortest paths problem. For both problems, our algorithm  provides bounds that improve upon the previous state-of-the-art in almost the entire range of parameters.
In particular, we provide an all-pairs shortest paths algorithm that requires $O(n^{5/3} \cdot \log^{2/3} n)$ time, even for $b = 1$, for all values of $D$.

We also devise the first algorithm with non-trivial complexity guarantees for computing exact shortest paths in the {\em multipass semi-streaming} model of computation.

From the technical viewpoint, our algorithm computes a hopset $G''$ of a skeleton graph $G'$ of $G$ {\em without first computing} $G'$ itself.
We then conduct a Bellman-Ford exploration in $G' \cup G''$, while computing the required edges of $G'$ {\em on the fly}. As a result, our algorithm computes {\em exactly} those edges of $G'$ that it really needs, rather than computing approximately the entire $G'$.
\end{abstract}


\thispagestyle{empty}
\newpage
\setcounter{page}{1}

\section{Introduction}
\label{sec:introd}

\subsection{Single-Source Shortest Paths}
\label{sec:intr_sssp}

We study the {\em distributed single-source shortest paths} (henceforth,  SSSP) problem in the $\CONGEST$ model of distributed computing. In this model, a communication network is modeled by a weighted undirected $n$-vertex graph $G = (V,E,\omega)$, $\omega(e) \ge 0$ for every edge $e \in E$, whose vertices host autonomous processors. The processors have distinct identity numbers (shortly, $\Id$s), typically\footnote{They can  be as well  from a larger range, i.e., at most polynomial in $n$.} from the range $\{1,\ldots,n\}$. The processors communicate via edges of the graph in synchronous rounds. In every round every vertex $v$ is allowed to send {\em short} messages to its neighbors. A message sent at the beginning of a round, arrives by the end of the same round. By ``short" one typically means $O(\log n)$ bits; alternatively, and somewhat more generally, one can also allow a message to contain up to $O(1)$ edge weights or/and vertex $\Id$s. In a yet  more general $\CONGEST(b \log n)$ model, for an integer parameter $b \ge 1$, 
one can deliver $O(b \log n)$ bits in each message, or more generally, $O(b)$ edges weights and/or vertex $\Id$s. When $b=1$, we write $\CONGEST$ for $\CONGEST(\log n)$.

The running time of an algorithm in this model is the (worst-case) number of rounds of distributed computation that it requires. At the beginning of an algorithm every vertex $v$ knows his $\Id$ number and the weights of edges incident on it. By the end of the algorithm, $v$ needs to know its {\em exact} distance  to a designated source vertex $r$, which is given as a part of the problem's input.\footnote{Vertices do not have to know $r$ at the beginning of the computation. It is sufficient that every vertex knows if it is the source $r$ or not.}  In the closely related {\em shortest path tree} (SPT) problem, $v$ also needs to know the identity of its parent $p(v)$ in the tree, and which edges $(v,u)$ among those incident on it belong to the SPT. 

The distributed SSSP problem is among the most central, extensively studied, and fundamentally important problems in this area. The classical Bellman-Ford algorithm \cite{B58,F56} has running time $O(n)$. There are instances, such as a weighted $n$-cycle $C_n$, in which the problem requires $\Omega(n)$ time, and thus the problem is said to be a {\em global} one, i.e., a problem for which one may need to traverse the entire network to solve it. (The notion of ``global" problem is due to Garay \etal \cite{GKP98}.) 
Moreover, Peleg and Rubinovich \cite{PR99} showed that the problem (as well as the MST problem) requires $\tOmega(D + \sqrt{n}/b)$ time;
this was later improved to $\tOmega(D + \sqrt{n/b})$ in \cite{E04}, where $D$ is the {\em hop-diameter} of $G$, i.e., the maximum unweighted distance between a pair of vertices $u$ and $v$ in $G$.

The results of \cite{B58,F56,PR99,E04} left open the gap between the upper bound of $O(n)$ and the lower bound of $\tOmega(D + \sqrt{n/b})$. Remarkably, the gap persists (and is actually even wider) if one restricts his attention to graphs with small diameter $D = O(1)$.
In this case the lower bound is $\tOmega(D + n^{1/2 - O(1/D)})$ \cite{E04,LPP06}. 
To the best of our knowledge, only for the case $D = 1$ (the so-called Congested Clique model) sublinear in $n$ bounds are known \cite{CHKK15}. 

Addressing this gap is one of the most central open problems in this area. The current author heard this question for the first time from Rubinovich some fifteen years ago; consequently, it was raised in the current author's survey \cite{E04:survey}.  (In the quotation below $B = b \log n$.)
\\
\\
``{\em What is the complexity of the shortest path tree problem in the message-passing model with small bandwidth parameter $B$ (i.e., $B = \log n$)?}"
\\
\\
This question was recently raised again in the Open Problems section of Nanongkai's paper \cite{N14}. 
\\
\\
``{\em Another question that should be very interesting is understanding the exact case: \\
Problem I.2 Can we solve SSSP} exactly {\em in sublinear time?}"
\\
\\
(The emphasis on ``exactly" is in \cite{N14}.) He then motivates the problem by saying:
\\
\\
``{\em In some settings, an exact algorithm for computing shortest paths is crucial; e.g., some Internet protocols such as OSPF and IS-IS use edge weights to control the traffic and using approximate shortest paths is unacceptable.}"
\\
\\
The same question was raised again in the Conclusions and Open Problems section of \cite{HKN15}. They wrote (the emphasis on ``approximately" and ``exactly" is in \cite{HKN15}):
\\
\\
``{\em Finally, while our paper essentially settles the running time for computing single-source
shortest paths approximately, the best running time for solving this problem exactly is
still linear (by the Bellman-Ford algorithm). Whether there is a sublinear algorithm is a
major open problem. In fact, in the past few years we have much better understood how to
{\em approximately} solve basic graph problems, such as minimum cut, single-source shortest paths,
all-pairs shortest paths, and maximum flows, on distributed networks (e.g. [NS14, GK13,
GKK+15]). However, when it comes to solving these problems {\em exactly}, almost nothing is
known. Understanding the complexity of exact algorithms is an important open problem."}
\\
\\

Lack of progress on this problem led researchers to consider its relaxed version, in which one is interested in {\em approximate} distances from a designated source $r$, rather than in the {\em exact} ones.  In STOC'13, Lenzen and Patt-Shamir \cite{LP13} devised an $O(k \log k)$-approximate algorithm for SSSP with running 
time $\tO(D + n^{1/2 +1/k})$. 
In STOC'14, Nanongkai came up with a $(1 + o(1))$-approximate algorithm that requires $\tO(n^{1/2} D^{1/4} + D)$ time. In STOC'16, Henzinger \etal \cite{HKN15} improved this bound further to $\tO(n^{1/2 + o(1)} + D)$ time. Finally, very recently, Becker \etal \cite{BKKL16} devised a $(1+\eps)$-approximate algorithm with running time $\tO(\eps^{-O(1)} (D + \sqrt{n}))$. We note that a lower bound of $\tOmega(D + \sqrt{n/b})$ \cite{E04,SHKKNPPW12} applies to approximate variants of the problem as well.

Nevertheless, despite this intensive research \cite{LP13,N14,HKN15,BKKL16,E04,SHKKNPPW12}, the fundamental question of whether {\em exact} SSSP can be solved in $o(n)$ time when the diameter $D$ is small remained wide open. In this paper we answer this question in the affirmative. Specifically, we devise a randomized algorithm that solves the problem in $O((n \log n)^{5/6})$ time  when $D = O(\sqrt{n \log n})$, and more generally, in $O(D^{1/3} \cdot (n \log n)^{2/3})$ time, for larger $D$. (The result applies to the $\CONGEST$ model, i.e., $B = \log n$.) Observe that this running time is sublinear in $n$ in {\em almost the entire range} of parameters, that is, as long as $D = o(n/\log^2 n)$. 
Moreover, our algorithm can compute an SPT rooted at a designated source $r$ within the same time. 

We note also that all previous sublinear-time $(1+\eps)$-approximate SSSP algorithms \cite{N14,HKN15,EN16,BKKL16} require time proportional to $\log \Lambda$, where $\Lambda = \frac{\max\{ \omega(e) \mid e \in E\}}{\min \{\omega(e) \mid e \in E\}}$ is 
the aspect ratio of the graph. This is, of course, unavoidable if the bandwidth is $O(\log n)$, because just delivering a single edge weight over an edge requires $O(\log_n \Lambda)$ time. However, in the natural model in which the bandwidth allows us to deliver a single edge weight or/and $\Id$ number through an edge in a round, the running time can be independent of $\Lambda$. 
This is the situation for a much-better-understood MST problem; the near-optimal algorithm of Kutten and Peleg \cite{KP98} requires $O(D + \sqrt{n}\log^* n)$ time in this model, even if the aspect ratio is huge. On the other hand, the state-of-the-art $(1+\eps)$-approximate SSSP algorithm \cite{BKKL16} requires $\tO(\eps^{-O(1)} (\sqrt{n} +D))\cdot  \log \Lambda$ time, i.e., it is sublinear in $n$ only if $\Lambda = 2^{o(\sqrt{n})}$. This is also the case with the approximate  algorithms from \cite{N14,HKN15,EN16}.\footnote{\cite{EN16} eliminate the dependence on $\Lambda$ from their SSSP algorithm for the Congested Clique and streaming models. However,  similarly to \cite{N14,HKN15},  the running time of their algorithm for the $\CONGEST$ model is proportional to $\log \Lambda$.}
 On the other hand, the running time of our exact SSSP algorithm, like the running time of the MST algorithm of Kutten and Peleg \cite{KP98}, is {\em truly sublinear}, i.e., in particular, independent of the aspect ratio $\Lambda$.

\subsection{Extensions and Applications}
\label{sec:ext}

\subsubsection{$s$ Sources}
\label{sec:many_sources}

We also extend our result in two directions. First, we consider the {\em exact} $s$-sources shortest paths (henceforth, $s$-SSP) problem, i.e., given a set $S$ of $|S| = s$ sources, we want to compute shortest paths for all pairs of vertices in $S \times V$. 

To the best of our knowledge, the only existing solution to this problem is to run the Bellman-Ford (henceforth, B-F) algorithm in parallel from all $s$ sources; due to congestion it requires $O(n \cdot s)$ time.  Our algorithm solves the $s$-SSP problem  (1) in time $O((n \log n)^{5/6} \cdot s^{2/3})$ for $D = O(s \sqrt{n \log n})$ and $s = O(\sqrt{n \log n})$; and (2) in time $O((n \log n)^{2/3} \cdot s)$, for $s = \Omega(\sqrt{n \log n})$ (applicable for all values of $D$). Together, these bounds improve the trivial $O(n \cdot s)$  bound of B-F for  the entire range of parameters. 
 In particular, our algorithm solves the {\em all-pairs shortest paths} (APSP) problem in time $O(n^{5/3} \cdot \log^{2/3} n)$, for all values of $D$. 

We remark that faster {\em approximate} $s$-SSP algorithms are known \cite{HKN15,EN16}; their running time is $\tO(D + \sqrt{n s})\cdot \log \Lambda$, for a sufficiently large $s$, and $\tO(D + \sqrt{ns}) \cdot n^{o(1)} \cdot \log \Lambda$ for all $s$. 
Also, Holzer and Wattenhofer \cite{HW12} devised an algorithm that solves {\em unweighted} APSP in $O(n)$ time. 

\subsubsection{Large Bandwidth}

We also extend our algorithm to work more efficiently when larger bandwidth $O(b \cdot \log n)$ is available, for $1 \le b \le n$, i.e., in the $\CONGEST(b \cdot \log n)$ model. (In fact, we assume that up to $b$ edge weights can be delivered through an edge in a single round.)

We are aware of only two (both of which are trivial) existing solutions for the {\em single-source} problem (SSSP) in this model. One is the B-F algorithm, which makes no use of larger bandwidth, and as a result requires $O(n)$ time.
The other one builds an auxiliary spanning BFS  tree  $\tau$ for $G$ rooted at an arbitrary vertex $\rt$. 
Then the trivial algorithm collects the entire topology of $G$ into $\rt$; solves the problem locally in $\rt$, and disseminates the solution. This algorithm requires $O(D + |E|/b)$ time.
Observe, however, that for dense graphs this expression  is also not sublinear in $n$. 

The results that we described in Section \ref{sec:intr_sssp} are sublinear even when the bandwidth is small. However, using larger bandwidth our algorithm solves the SSSP problem in the $\CONGEST(b \cdot \log n)$ model (1) in time $O\left({{(n \log n)^{5/6}} \over {b^{1/2}}}\right)$, if $b \le (n \log n)^{1/3}$, and $D = O\left(\sqrt{{n \log n} \over b}\right)$; and (2) in time $O\left({{(n \log n)^{3/4}} \over {b^{1/4}}}\right)$, if $b \ge (n \log n)^{1/3}$, $D = O\left(\sqrt{{n \log n} \over b}\right)$; and (3) in time $O((D/b)^{1/3} (n \log n)^{2/3})$, for larger values of $D$, under certain mild restrictions on $D$ and $b$ (see Theorem \ref{thm:single_src_large_band} for details).

Note that bound (2) above gives running time $\tO(\sqrt{n})$, when the bandwidth $b$ is really large (i.e., $b \approx n/\polylog(n)$), and the diameter $D$ is rather small ($D \le \polylog(n)$). We remark, however, that the lower bound in $\CONGEST(b \cdot \log n)$ model in this regime behaves like $\tOmega\left(\sqrt{n \over b}\right)$ \cite{E04}; i.e., polylogarithmic-time SSSP might be possible for such a large bandwidth and small diameter.

Finally, we also extend our algorithm to the $s$-SSP problem in the $\CONGEST(b \cdot \log n)$ model. In this  case the B-F algorithm can be sped up, and it requires $O(n \lceil s/b\rceil)$ time. The topology-collecting algorithm, described above, requires $O\left(D + {{|E| + ns} \over b}\right)$ time. Our results for this setting generalize our single-source and/or unit-bandwidth results, and improve the existing (trivial) bounds in almost the entire range of parameters. (See Theorem \ref{thm:large_band_largeD}, and the discussion that follows it, for details.) 

\subsubsection{Streaming Model}
\label{sec:introd_stream}

A variant of our algorithm provides also the first non-trivial upper bound for the SSSP  problems in the multipass {\em semi-streaming} model. In this model, the algorithm is allowed to read the stream of edges of the input $n$-vertex graph $G = (V,E)$ multiple times (aka passes), while storing only a limited amount of information at all times. 

The problem of computing SSSP  (with respect to a designated root vertex $r$) using a small number of passes and small memory is one of the most central open questions in the area of semi-streaming\footnote{When the allowed memory is $\Omega(n)$, the model is typically called ``semi-streaming", as opposed to the ``strict" streaming model, in which the allowed space is $o(n)$. Since most graph problems require $\Omega(n)$ space, they are typically studied in the semi-streaming model \cite{FK+04,FK+05}.} graph algorithms. Feigenbaum et al. \cite{FK+04,FK+05} pioneered the study of graph problems in this model. They observed that the greedy algorithm for constructing graph spanners \cite{ADDJ90} provides a single-pass streaming approximate algorithm for computing graph distances, and devised a more efficient (in terms of processing time-per-edge) algorithm for this problem. Their results were consequently improved in \cite{E11,B06b}. Following this research direction, \cite{EZ06,EN17} devised efficient streaming multipass algorithms for computing sparse $(1+\eps,\beta)$-spanners for unweighted graphs, and as a result, derived improved multipass streaming algorithms for computing approximate distances and paths.
Ahn et al. \cite{AGM12} devised efficient algorithms for these problems in the so-called {\em turnstile} (aka {\em dynamic}) multipass semi-streaming model. See \cite{AGM12} for the definition of this model.

On the lower bound frontier, Feigenbaum et al. \cite{FK+04,FK+05} showed that even in {\em unweighted} undirected graphs, computing a $p$-neighborhood of a given vertex , for a positive integer parameter $p = O\left({{\log n} \over {\log \log n}}\right)$, using $p-1$ passes or less, requires $n^{1+\Omega(1/p)}$ space. Guruswami and Onak \cite{GO16} showed that nearly the same lower bound (up to factors polynomial in $\log n$ and in $p$) applies to an easier problem of computing an exact distance  between a given pair of vertices, which are at distance $\Theta(p)$ from one another.

Note, however, that no non-trivial {\em upper} bounds are known for the fundamental problem of {\em exact} SSSP computation, even in {\em unweighted} graphs. 
This problem appears explicitly in the collection of open problems from Kanpur's Workshop on Algorithms for Data Streams, 2006, compiled by McGregor \cite{McG06}. Specifically, problem 14 in this collection says:
\\
\\
{\em ``Clearly, $d_G(u,v)$ can be computed exactly in $d_G(u,v)$ passes, but for large $d_G(u,v)$ this is infeasible. Can we do better?"}
\\
\\
In other words, there are two trivial upper bounds for exact SSSP in this model. The first one  (outlined in McGregor's question) uses $O(n)$ passes and $O(n)$ memory, and the second one (which stores the entire graph) uses just one single pass, but $O(|E|)$ memory.

In the current paper we devise a deterministic algorithm that smoothly interpolates between these two trivial bounds. This algorithm, for a parameter $k$, $1 \le k \le n$, computes SSSP in weighted undirected graphs with non-negative edge weights, in $O(n/k)$ passes over the stream and $O(n k)$ memory. We also generalize this result to the $s$-SSP problem, and show that for any parameter $k \ge s$, our algorithm solves $s$-SSP using the same number of passes and memory as in the single-source case. (See Section \ref{sec:stream} for further details.)
 
Moreover, using randomization, we extend this result to directed graphs with possibly negative edge weights. However, the number of passes in this generalized result becomes larger by a logarithmic factor, i.e., it becomes $O((n/k) \log n)$. 

\subsubsection{Approximating Diameter}
\label{sec:diam}

Once single-source distances from a designated root vertex $r$ are computed, it is easy to compute the radius with respect to $r$ (i.e., the maximum distance between some vertex $v \in V$ and $r$) within additional $O(D)$ distributed time. Observe that the (weighted) {\em diameter} (i.e., the maximum distance between some pair $u,v \in V$ of vertices) is at least  the radius, and is at most twice the radius.
Hence our algorithm also provides a 2-approximation of the (weighted) diameter for the input graph, within the same time bounds.
Previous sublinear-time algorithms for approximating weighted diameter \cite{N14,HKN15} provide $(2 + o(1))$-approximation.
On the lower bound frontier, Henzinger et al. \cite{HKN15}, citing \cite{HW12}, state that a $(2-\eps)$-approximation of weighted diameter, for any constant arbitrarily small $\eps >0$, requires $\tOmega(n)$ distributed time.
The problem of estimating diameter in distributed setting was extensively studied; cf. \cite{FHW12,HW12,HPRW14,HP15}, and the references therein. 

\subsection{Technical Overview}
\label{sec:intr_tech}

The basic approach in many recent distributed approximate shortest paths algorithms \cite{LP13,N14,HKN15,EN16} is the following one: one samples\footnote{\cite{HKN15} replaced this sampling by a deterministic selection. Also, the exact size of $V'$ may vary between different algorithms, and depend on problem's parameters, such as the diameter, the bandwidth and the number of sources.} roughly $\sqrt{n}$ ``virtual" vertices. Denote the set of virtual vertices by $V'$. Then the algorithm builds a virtual or ``skeleton" graph $G' = (V',E',\omega')$ on the set $V'$ of virtual vertices. A pair $(u',v')$ of virtual vertices forms an edge in $E'$ if there exists a path $\pi(u',v')$ between them in $G$  with at most $c \cdot \sqrt{n} \cdot \ln n$ hops, for an appropriate constant $c$. If it is the case, the weight $\omega'(u',v')$ of the edge $(u',v') \in E'$ 
is the weight of the shortest
 such a $(c \cdot \sqrt{n} \cdot \ln n)$-limited $u'-v'$ path in $G$. (A path is said to be {\em $h$-limited}, for a parameter $h$, if it contains at most $h$ hops. The smallest weight of an $h$-limited $u'-v'$ path in $G$ is called the {\em $h$-limited distance} between $u'$ and $v'$, and is denoted $d_G^{(h)}(u',v')$.) 

Once $G'$ is constructed, the algorithm builds a {\em hopset} $G''$ for $G'$. A graph $G'' = (V',H',\omega'')$ is said to be a {\em $(\beta,\eps)$-hopset} for $G'$, for an integer parameter $\beta > 0$ and a real parameter $\eps > 0$, if for every pair $u',v' \in V'$ of vertices in $G'$, we have
$$d_{G'}(u',v') ~\le~ d_{G' \cup G''}^{(\beta)}(u',v') ~\le~ (1 + \eps) d_{G'}(u',v')~.$$
Here $G' \cup G'' = (V', E' \cup H',\hat{\omega})$, where the weight function $\hat{\omega}$ gives preference to hopset edges, i.e., for $e \in H'$, 
we have $\hat{\omega}(e) = \omega''(e)$, and for $e \in E' \setmns H'$, we have $\hat{\omega}(e) = \omega'(e)$.
Efficient distributed constructions of sparse hopsets with $\beta = n^{o(1)}$ can be found in \cite{HKN15,EN16}. The main precursor of these constructions  is the PRAM construction of hopsets from \cite{C00}.

After the graph $G'$ on the vertex set $V'$ and the hopset $G''$ for $G'$ are constructed, the algorithm conducts a B-F exploration in $G' \cup G''$ originated from the designated source vertex $r$. (It can be assumed that $r \in V'$.) In each iteration of this B-F exploration, every vertex $v'$ from $V'$ broadcasts its current distance estimate (which is an upper bound on $d_{G'}(r,v')$) to the entire graph. The number of iterations of this B-F exploration is at most the hopbound $\beta$ of the hopset $G''$. Since $\beta = n^{o(1)}$, the entire algorithm is efficient.

When one tries to use this approach for computing {\em exact} shortest paths, the main hurdle is that it is not even clear how to compute the virtual graph $G'$. All existing algorithms \cite{N14,HKN15,EN16} rely on {\em approximate} computation of paths with limited number of hops.
Specifically, Nanongkai \cite{N14} developed an elegant and sophisticated routine which computes $h$-limited $(1+\eps)$-approximate shortest paths from $s$ designated sources in $\tO(D + h + s) \cdot \log \Lambda$  time. In the context of the single-source problem, it holds that $h \approx s \approx \sqrt{n}$, and therefore an {\em approximate} version of the virtual graph $G'$ can be computed efficiently, i.e., in $\tO(D + \sqrt{n}) \cdot \log \Lambda$ time. However, obviously, once $G'$ is computed approximately, the entire scheme is doomed to compute approximate distances, even if one were using an exact hopset $G''$ of $G'$.

Our main idea is to {\em bypass} the computation of $G'$. We show that, perhaps surprisingly, one can compute a hopset $G''$ for $G'$ {\em without} computing the virtual graph $G'$ first! Once this is done, we still need to conduct a B-F exploration in $G' \cup G''$, and a-priori this also seems to require the vertices to know $G'$. We observe, however, that the entire graph $G'$ is not needed. Rather one can compute only those edges of $G'$ that the B-F exploration in $G' \cup G''$ traverses. Our algorithm does this {\em on the fly}, i.e., during the exploration. Albeit, these edges are computed {\em exactly}, rather than approximately. 

Next, we sketch how our algorithm constructs  a hopset $G''$ without computing the virtual graph $G'$ first. Naturally, the hopset $G''$ needs to be {\em exact}, i.e., with $\eps = 0$, as we aim at exact distances. Exact hopsets were built in \cite{UY91,KS97,SS99}.  We use the hopset of Shi and Spencer \cite{SS99}, called the {\em $k$-shortcut hopset}, for a parameter $k \ge 1$. The hopset ${G'}^{(k)}$ for $G' = (V',E',\omega')$ contains, for every vertex $v' \in V'$, edges $(v',v'_1), \ldots,(v',v'_k)$ connecting $v'$ to the $k$ closest vertices $v'_1,\ldots,v'_k \in V'$ to $v'$.\footnote{
The observation that the structure that \cite{SS99} is a hopset with hopbound $O(|V'|/k)$ appears explicitly in \cite{C00}. Nanongkai \cite{N14} provided an explicit analysis and a distributed construction  of this hopset, given an approximate virtual graph $G'$.}
We show that an exact $k$-shortcut hopset ${G'}^{(k)}$ can be constructed without knowing $G'$ in $\tO(\sqrt{n} \cdot k^2)$ time. 

The algorithm  that builds this hopset is actually very simple. We run a B-F exploration in the original graph $G$ from virtual vertices $v' \in V'$ in parallel to depth $\tO(\sqrt{n} \cdot k)$. In each iteration of this exploration, every vertex $v \in V$ records and forwards only the $k$ closest virtual vertices that it knows. As a result, due to congestion, every iteration lasts for $O(k)$ rounds. At the end of the exploration, every vertex $v \in V$ knows the $k$ closest virtual vertices to it (with respect to distance in $G$) that are reachable from $v$ via $\tO(\sqrt{n} \cdot k)$-limited paths from the original graph $G$. It is not hard to see that, whp, these are exactly the $k$ closest to $v$ virtual vertices {\em with respect to distance in $G'$}, i.e., this exploration actually computes the desired $k$-shortcut hopset. 

Once the hopset $G''$ is computed, we run a B-F exploration in $G' \cup G''$ originated at $r$ for $\beta = O(|V'|/k)$ iterations.
Each iteration involves computing the required edges of $G'$, and updating distance estimates via the edges of $G'$ that were just computed and via the hopset edges of $G''$. The former is done by an inner B-F exploration in $G$, and the latter is done via a broadcast in the auxiliary BFS tree $\tau$ of the entire graph. For concreteness, consider the special (but a very important) case of single source and small diameter. Then
each edge of $G'$ corresponds to a path of length $\approx \sqrt{n}$ in $G$, and thus the B-F exploration in $G$ goes to depth $\approx \sqrt{n}$. The broadcast in the BFS tree $\tau$ requires $O(D + |V'|) \approx \sqrt{n}$ time. Thus, a single  iteration of this B-F exploration in $G' \cup G''$ is implemented in $\approx \sqrt{n}$ time. Since there are $\beta=O(|V'|/k) \approx \sqrt{n}/k$ iterations, this entire exploration requires $\approx n/k$ time. This expression is balanced with the time required to construct the hopset $G''$ (roughly $\sqrt{n} \cdot k^2$); this yields time $\approx n^{5/6}$.

We note that our entire algorithm is pretty simple, and it involves no heavy local computations. The algorithm amounts to a number of carefully combined B-F explorations. In particular, it is much simpler than the existing approximate SSSP solutions \cite{N14,HKN15,EN16,BKKL16}. 
Indeed, \cite{N14,HKN15,EN16} all involve sophisticated ``light-weight" approximate shortest path computations to compute $G'$, and then employ intricate constructions of approximate hopsets. The recent algorithm of \cite{BKKL16} bypasses  hopsets altogether. It, however, employs a sophisticated and technically very involved constrained gradient descent method. 

\subsection{Related Work}
\label{sec:rel_work}

To the best of our knowledge, the research thread that aims at solving global distributed problems in sublinear in $n$ time when the diameter $D$ is relatively small was initiated by Peleg \cite{P90} in the context of Leader Election problem. Awerbuch \cite{A89}, citing an unpublished manuscript by Peleg, says: ``{\em Peleg points out that the difference between $O(V)$ and $O(D)$ time can be very significant in many existing networks, e.g. the ARPANET, where $D \ll V$.}" 

Bellman-Ford algorithm \cite{B58,F56} was developed a few decades before  before the distributed message-passing model was formalized.
Explicit descriptions of distributed Bellman-Ford algorithm were given, e.g., by Gallager in \cite{Gal76,Gal82,BG92}.

In addition to the SSSP and MST, other global problems for which sublinear in $n$ time algorithms were devised include the asynchronous BFS \cite{A89}, and the minimum cut problem \cite{GK13,NS14}. Distributed exact shortest paths algorithms were also devised in \cite{J85,F85,A89}.


The idea of computing a spanner on a virtual graph without first computing the virtual graph itself appeared in \cite{LP13}. Note, however, that once a spanner is constructed, one no longer needs the virtual graph, but rather can conduct distance computations in the spanner itself.
This is not the case with a hopset. Even when the hopset $G''$ of the virtual graph $G'$ is already constructed, one still needs the graph $G'$ to estimate distances. (Recall that distance computations are then conducted in $G' \cup G''$.) The observation that one can conduct all these computations without ever computing the virtual graph $G'$ itself  is crucial for our result. 

\subsection{Structure of the Paper}
In Section \ref{sec:hopset} we describe our algorithm that constructs  a $k$-shortcut hopset for the skeleton graph $G'$. 
In Section \ref{sec:dist} we show how the hopset can be used to compute single-source distances in sublinear time.
For simplicity of presentation, the bounds derived in Sections \ref{sec:hopset} and \ref{sec:dist} are not the best ones that we can get.
We derive sharper bounds  in Sections \ref{sec:impr} and \ref{sec:mult_src}.
Our streaming algorithm and its analysis are provided in Section \ref{sec:stream}.
 In Appendix \ref{sec:large_band} we adapt our algorithms to the scenario that the bandwidth parameter $b$ is larger than one. Some proofs are deferred to Appendix \ref{app:pf}. Finally, In Appendix \ref{sec:upcast} we generalize a classical upcast algorithm to $\CONGEST(b\log n)$ model.

\section{Computing the $k$-Shortcut Hopset}
\label{sec:hopset}

Consider a weighted undirected graph $G = (V,E,\omega)$, and let $k$ be a positive integer parameter.
For a vertex $v \in V$, let $S_G[k](v)$ (or, shortly, $S[k](v)$, when $G$ can be understood from the context) denote  the set of $k$ closest 
(reachable) vertices to $G$, not including $v$. Ties are broken arbitrarily.  We define the following graph $G^{(k)} = (V,H,\omega^{(k)})$ on top of $G$.
$$
H = \{(u,v) \mid u \in S_G[k](v) \mbox{~or~} v \in S_G[k](u)\}~.
$$
The weight function $\omega^{(k)}$ is defined by $\omega^{(k)}(u,v) = d_G(u,v)$.
The graph $G^{(k)}$ will be referred to as the {\em $k$-shortcut hopset} of $G$. 
The following theorem (from \cite{SS99,C00,N14}) justifies the name of $G^{(k)}$. We provide the proof of this theorem, which follows closely the proof from \cite{N14}, in Appendix \ref{app:pf} for the sake of completeness.

\begin{theorem} \cite{SS99,C00,N14}
\label{thm:hopset}
$G^{(k)}$ is an exact hopset of $G$ with hopbound $h = O(n/k)$. 
\end{theorem}

We sample $N = \Theta(\sqrt{n} \log n)$ vertices $V' = \{v_1,v_2,\ldots,v_N\}$, i.e., every vertex is sampled independently at random with probability $q = {{c \ln n} \over {\sqrt{n}-1}}$, for a sufficiently large constant $c$. By Chernoff's bound, $N = \Theta(\sqrt{n}\log n)$.
Consider the virtual graph $G' = (V',E')$, where
$$
E' = \{(v',u') \mid v',u' \in V', \mbox{~} \exists v'-u' \mbox{~path~in~} G \mbox{~with~} \le \sqrt{n} \mbox{~hops}\}.
$$
The weight function $\omega'$ is defined by $\omega'(v',u') = d_G^{(\sqrt{n})}(v',u')$.

Our first objective is to compute a $k$-shortcut hopset  ${G'}^{(k)}$ of $G'$, for a parameter $k$, {\em without} first computing the virtual graph $G'$.  To this end we employ the following variant of the B-F  algorithm.

We divide the computation into {\em super-rounds}, each lasting for $O(k)$ rounds. At the beginning of each super-round $i$,
$i = 0,1,\ldots$, every vertex $v$ maintains the set $\cS^{(i)}(v) = {S'}_G^{(i)}[k](v)$, defined as the set of $k$ closest selected (aka, virtual) vertices $v' \in V'$ to $v$, closest with respect to $i$-limited distance in $G$. In other words, if one orders virtual vertices $v'_1,\ldots,v'_N$ in the order of non-decreasing $i$-limited distance from $v$ in $G$ (i.e., $d_G^{(i)}(v,v'_1) \le d_G^{(i)}(v,v'_2) \le \ldots \le d_G^{(i)}(v,v'_N)$), then  
$\cS^{(i)}(v)$ contains the first $k$ elements in this ordering.  
(It is also required that $d_G^{(i)}(v,v'_j) < \infty$, for any $v'_j \in \cS^{(i)}(v)$.)

Here ties are broken in the following way. For $v'_h,v'_j$, $1 \le h < j \le N$, suppose that $d_G^{(i)}(v,v'_h) = d_G^{(i)}(v,v'_j)$.
Let $h'$ (respectively, $j'$) be the number of hops in the shortest $i$-limited $v-v'_h$ (resp., $v-v'_j$) path $\pi(v,v'_h)$ (resp., $\pi(v,v'_j)$) in $G$. Then if $h' < j'$, then we write $v'_h <_v v'_j$, i.e.,  we prefer $v'_h$ to $v'_j$ in the ordering $<_v$. Symmetrically, if $h' > j'$, then $v'_h >_v v'_j$. Finally, if $h'=j'$, then the paths $\pi(v,v'_h), \pi(v,v'_j)$ are compared in the lexicographical order, starting with $v$. (We assume that all vertices have distinct $\Id$ numbers.) We write $\pi(v,v'_h) <_v \pi(v,v'_j)$, if $\pi(v,v'_h)$ is lexicographically smaller than $\pi(v,v'_j)$, and then also $v'_h <_v v'_j$.

This completes the definition of $\cS^{(i)}(v)$. Recall that we assume inductively that at the beginning of a super-round $i$, $i = 0,1,\ldots$, every vertex $v$ knows $\cS^{(i)}(v)$. For every $v' \in \cS^{(i)}(v)$, the vertex $v$ also keeps the $i$-limited distance $d_G^{(i)}(v',v)$, the number of hops $h^{(i)}(v',v)$ in the shortest $i$-limited $v'-v$ path, and the {\em parent} or {\em predecessor} $u = p(v)$ from which $v$ received this tuple. Observe that the induction base case $i=0$ holds. 

In super-round $i$, every vertex $v$ sends to all its neighbors its entire set $\cS^{(i)}(v) = {S'}_G^{(i)}[k](v)$. Then $v$ selects $k$ smallest estimates among those that it received, using appropriate (see below) tie-breaking rules. As a result, the vertex $v$ computes its set $\cS^{(i+1)}(v)$.

Specifically, $v$ receives from a neighbor $u$ a tuple $(x,d^{(i)}(x,u),h^{(i)}(x,u),p(u))$. Then $v$ computes its own tuple
$(x,\delta_u^{(i+1)}(x,v) = d^{(i)}(x,u) + \omega(u,v),h_u^{(i+1)}(x,v) = h^{(i)}(x,u)+1,p(v) = u)$. 
(Here $\delta_u^{(i+1)}(x,v)$ is the estimate of $(i+1)$-limited $x-v$ distance that $v$ learns from $u$, and $h^{(i+1)}_u(x,v)$ is the number of hops in this esimate.)

Then for each origin $x$  that $v$ hears from, it computes the estimate $d^{(i+1)}(x,v) = \min_u \delta_u^{(i+1)}(x,v)$.
In case of equality, $v$ keeps the tuple with smaller $h_u^{(i+1)}(x,v)$. If these are equal too, then it prefers the tuple with lexicographically smaller parent $u$.
When comparing the tuples from different origins, $v$ uses the same rules again. 

Next, we argue that $\cS^{(i+1)}(v) = {S'}_G^{(i+1)}[k](v)$. For convenience, super-rounds are numbered below starting with 0.
The proof of this lemma is in Appendix \ref{app:pf}.

\begin{lemma}
\label{lm:hop_constr}
The set  $\cS^{(i+1)}(v)$ that each vertex $v$ computes at the end of super-round $i$ is equal to $ {S'}_G^{(i+1)}[k](v)$.
\end{lemma}
\def\Pfhopconstr
{
Let  $\Gamma(v) = \{u_1,u_2,\ldots,u_d\}$, $d = \deg(v)$,  denote the neighborhood of $v$ (in $G$). In super-round $i$, the vertex $v$ learns from its neighors the sets $\StagGi(u_1),\ldots,\StagGi(u_d)$. Consider a tuple $\sigma = (x,d^{(i+1)}(x,v),h^{(i+1)}(x,v),u) \in \StagGiplus(v)$, $u \in \Gamma(v)$,  and let $\pi(x,v)$ be the shortest $(i+1)$-limited $x-v$ path, $\omega(\pi(x,v)) = d^{(i+1)}(x,v)$, $|\pi(x,v)| = h^{(i+1)}(x,v)$. 

We argue that this tuple is in $\cS^{(i+1)}(v)$ as well. 
To this end, we first show that the tuple $\sigma_u = (x,d^{(i)}(x,u),h^{(i)}(x,u),p(u))$, where $p(u)$ is the predecessor of $u$ in $\pi(x,v)$, must belong to $\cS^{(i)}(u) = \StagGi(u)$. 

Indeed, suppose for contradiction that 
$$\StagGi(u) = \{(x_1,d^{(i)}(x_1,u),\hi(x_1,u),p_1(u)),\ldots,(x_k,\di(x_k,u),\hi(x_k,u),p_k(u))\}$$
 does not contain the tuple $\sigma_u$, where for every $j$, $1 \le j \le k$, $p_j(u)$ is the predecessor of $u$ in $\pi(x_j,u)$. 

Then, for each $j$, $1 \le j \le k$, either $\di(x_j,u) < \di(x,u)$, or $\di(x_j,u) = \di(x,u)$ but $\hi(x_j,u) < \hi(x,u)$, or both 
$\di(x_j,u) = \di(x,u)$ and $\hi(x_j,u) =\hi(x,u)$, but the path $\pi(u,x_j)$ is lexicographically smaller than $\pi(u,x)$, i.e., 
$\pi(u,x_j) <_u \pi(u,x)$. 

Denote $\pi(x_j,v) = \pi(x_j,u) \circ (u,v)$, for every $1 \le j \le k$. (Here $\circ$ stands for concatenation.) It follows that for each $j$, $1 \le j \le k$, the tuple $(x_j,d^{(i+1)}(x_j,v),h^{(i+1)}(x_j,v),u)$ is better than $(x,d^{(i+1)}(x,v),h^{(i+1)}(x,v),u)$ (with respect to $<_u$).
This is a contradiction to the assumption that $\sigma \in \StagGiplus(v)$. 

Hence $\sigma_u = (x,\di(x,u),\hi(x,u),p(u)) \in \StagGi(u) = \cS^{(i)}(u)$, and so $v$ receives this tuple from $u$ on super-round $i$. As a result, the vertex $v$ computes the tuple $\hsigma = (x,\deltaiplus_u(x,v),\hiplus_u(x,v),u)$, with $(\deltaiplus_u(x,v),\hiplus_u(x,v)) = 
(\diplus(x,v),\hiplus(x,v))$.  (Note that  $\hsigma = \sigma$.) 

Consider a comparison that $v$ conducts between $\sigma= \hsigma$ and some other tuple 
$$\sigma' = (x',\diplus(x',v),\hiplus(x',v),p'(v)) =  \hsigma' = (x',\deltaiplus_{p'(v)}(x',v),\hiplus_{p'(v)}(x',v),p'(v)),$$ 
where $p'(v)$ is the parent of $v$ in this tuple. (The vertex $v$ has  computed $\hsigma'$ in this super-round.)  Assume that $\sigma <_v \sigma'$, i.e., that $(\diplus(x,v),\hiplus(x,v)) < (\diplus(x',v),\hiplus(x',v))$, or the two pairs are equal, but $\pi(x,v) <_v \pi(x',v)$, where $\pi(x,v)$ (respectively, $\pi(x',v)$) is the shortest $(i+1)$-limited $x-v$ (resp., $x'-v$) path with the smallest number of hops. 

If $x = x'$, then $v$ received this tuple from two distinct neighbors $u$ and $u'$, and it prefers $\sigma$ because either
 $(\deltaiplus_u(x,v),\hiplus_u(x,v)) < (\deltaiplus_{u'}(x,v),\hiplus_{u'}(x,v)) $, or the two pairs are equal but $\Id(u) < \Id(u')$.
(Otherwise there were a contradiction to $\sigma <_v \sigma'$.)

So we assume that $x \neq x'$. If $(\diplus(x,v),\hiplus(x,v)) < (\diplus(x',v),\hiplus(x',v))$, then obviously $v$ prefers $\sigma$ over $\sigma'$.
So it remains only to consider the case that the two pairs are equal. Denote by $u = p(v)$ and $u' = p'(v)$ the predecessors of $v$ in $\pi(x,v)$ and $\pi(x',v)$, respectively. If $u \neq u'$ then $\Id(u) < Id(u')$, and so $v$ prefers $\sigma$ over $\sigma'$ (as it knows both $u$ and $u'$).
Finally, consider the case that $u = u'$. Then $u$ has sent both the tuples $\sigma_u$ and $\sigma'_u$ to $v$, with an indication that $\sigma_u <_u \sigma'_u$. (The tuples $\sigma$ and $\sigma'$ are computed by $v$ from the tuples $\sigma_u$ and $\sigma'_u$, respectively.) So $v$ concludes that $\sigma <_v \sigma'$ in this case too.

To summarize, based on the information that $v$ receives, it can compare the tuples correctly. Since it receives all the tuples of $\StagGiplus(v)$ (along with possibly some other tuples), it therefore computes correctly the set of $k$ smallest tuples with respect to $<_v$, i.e., the set $\cS^{(i+1)}(v)$ that it computes is equal to $\StagGiplus(v)$.
\QED
}

To summarize, after $\sqrt{n} \cdot k$ super-rounds of this Bellman-Ford algorithm (which last for $O(\sqrt{n} \cdot k^2)$ time), every vertex $v$ computes ${S'}_G^{(\sqrt{n} \cdot k)}[k](v)$.\footnote{As pointed out by an anonymous reviewer of STOC'17 conference, this fact can also be derived using the so-called ``short-range scheme''  of \cite{LP13}.}

Next, we define the sets $\Stagtagi(v)$, for vertices $v' \in V'$, for some fixed integer $i$.
For a vertex $v' \in V'$, the set $\Stagtagi(v')$ is defined as the set of $k$ closest sampled vertices (i.e., vertices of $V'$) to $v'$ with respect to $i$-limited distances in $G'$. (The set will only contain vertices reachable within $i$ hops in $G'$ from $v'$. An analogous restriction applies also to the set $\StagGi(v')$.) In other words, let
$$\di_{G'}(v',v'_1) \le \di_{G'}(v',v'_2) \le \ldots \di_{G'}(v',v'_N)~.$$
Then $\Stagtagi(v') = \{v'_1,v'_2,\ldots,v'_k\}$, where the ties are broken according to the paths in $G$ (and not in $G'$).

Specifically, consider a pair of vertices $u',w'$ such that $\di_{G'}(v',u') = \di_{G'}(v',w')$.
Let $\pi(v',u')$ (respectively, $\pi(v',w')$) be the shortest $(i\cdot \sqrt{n})$-limited $v'-u'$ (resp., $v'-w'$) path in the {\em original} graph  $G$ with the smallest number of hops. Then $u' <_{v'} w'$ iff $\pi(v',u) <_{v'} \pi(v',w')$. 

Our next objective is to show that, with high probability (henceforth, we will write ``whp"), $\Stagtagk(v') = \Stagsqrt(v')$.
Before proving it, we define a collection of $i$-limited paths, for all $i \in [n-1]$, between all pairs of vertices, and argue that this collection admits useful nesting properties.

Let $\cP = \{P^{(i)}(u,v) \mid 1 \le i \le n-1, u \neq v, u,v \in V\}$ be the collection of $i$-limited shortest paths computed if we were to run B-F on $G$, with the above rules to break ties, from all vertices of the graph, one after another. (Note that our tie-breaking rules prefer paths that are lexicographically smaller when viewing them from the {\em tail to head}.)  We remark that the algorithm does not actually do this computation. We just imagine that it is done for  definitional purposes. 

The proofs of the following two lemmas are  in Appendix \ref{app:pf}.

\begin{lemma}
\label{lm:nest}
For any two paths $P^{(i)}(u,v), {P'}^{(i')}(u',v') \in \cP$ (recall that the paths are $i$-limited and $i'$-limited, respectively) that traverse some pair of vertices $x$ and $y$ in the same order, and the number of hops $\ell$ in $P^{(i)}(u,v)$ between $x$ and $y$ is equal to the number of hops in ${P'}^{(i')}(u',v')$ between $x$ and $y$, then the entire subpaths $P^{(\ell)}(x,y)$ and ${P'}^{(\ell)}(x,y)$ of the two respective paths are equal. In particular, it follows that $P^{(\ell)}(x,y) = {P'}^{(\ell)}(x,y) \in \cP$. 
\end{lemma}
\def\Pfnest
{
If $\omega(P^{(\ell)}(x,y)) < \omega({P'}^{(\ell)}(x,y))$, then ${P'}^{(i')}(u',v')$ is not the shortest $i'$-bounded $u'-v'$ path, contradiction.
Hence $\omega(P^{(\ell)}(x,y)) = \omega({P'}^{(\ell)}(x,y))$.

Write $P^{(\ell)}(x,y) = (x = x_0,x_1,\ldots,x_\ell = y)$, ${P'}^{(\ell)}(x,y) = (x =x'_0,x'_1,\ldots,x'_\ell = y)$. 
Let $h < \ell$ be the largest index such that $x_h \neq x'_h$. (Higher-index vertices coincide.) Assume without loss of generality (henceforth, we will write ``wlog") that $\Id(x_h) < \Id(x'_h)$.  Replace ${P'}^{(\ell)}(x,y)$ in ${P'}^{(i')}(u',v')$ by $P^{(\ell)}(x,y)$. We obtain a shortest $i$-limited $u'-v'$ path with weight at most $\omega({P'}^{(i')}(u,v))$, but which is lexicographically smaller. This is a contradiction to the assumption that ${P'}^{(i')}(u,v) \in \cP$. (Recall that $\cP$ was obtained via B-F explorations that return lexicographically smallest paths, when considering them from tail to head.)
\QED
}

\begin{lemma}
\label{lm:sqrt}
Whp, (specifically, with probability at least $1- n^{-(c-3)}$), any path $P^{(i)}(u,v) \in \cP$ with $|P^{(i)}(u,v)| \ge \sqrt{n}$ (i.e., the number of hops in the path is at least $\sqrt{n}$), contains at least one internal vertex from $V'$.
\end{lemma}
\def\Pfsqrt
{
For a single path $P$, $|P| \ge \sqrt{n}$, the probability for it not to contain a vertex of $V'$ as an internal vertex is at most 
$$ \left(1 - {{c \ln n} \over {\sqrt{n}-1}}\right)^{|V(P)|-2} ~=~ 
\left(1 - {{c \ln n} \over {\sqrt{n} -1}}\right)^{|P|-1} 
~\le~  \left(1 - {{c \ln n} \over {\sqrt{n}-1}}\right)^{\sqrt{n}-1} ~\le n^{-c}~.$$
By union-bound, the probability that some $P \in \cP$ with $|P| \ge \sqrt{n}$ not to contain an internal selected vertex is at most $n^{-(c-3)}$. Hence the assertion of the lemma holds with probability at least $1 - n^{-(c-3)}$.
\QED
}

Denote by $\cA$ the event of Lemma \ref{lm:sqrt}. We showed that $\Prob(\cA) \ge 1 - n^{-(c-3)}$. 

\begin{lemma}
\label{lm:path}
Conditioned on $\cA$, for every virtual vertex $v' \in V'$, and every $u' \in \Stagsqrt(v')$, there is a $k$-limited path in $G'$ of weight $d_G^{(\sqrt{n}k)}(u',v')$. 
\end{lemma}
\proof
Consider $P^{(\sqrt{n}k)}(v',u')$, i.e., the shortest $v'-u'$ $(\sqrt{n}k)$-limited path in $G$ with the fewest number of hops, smallest with respect to $<_{v'}$ among such paths.
(If there is no such a path, then $d_G^{(\sqrt{n}k)}(v',u') = \infty$, and there always exists a weight-$\infty$ (or less) $k$-limited path in $G'$ between $v'$ and $u'$.)

Let $v' = v'_0,v'_1,\ldots,v'_{h-1},v'_h=u'$ be the distinct selected vertices in $P^{\sqrk}(v',u')$. If $h > k$ (i.e., $h-1 \ge k$), 
then $v'_1,v'_2,\ldots,v'_{h-1}$
are all closer  to $v'$ than $u'$ with respect to $\sqrk$-bounded distance in $G$. (Or, if zero weights are allowed, then some of these vertices may be at the same distance from $v'$ as $u'$, but the number of hops between $v'$ and them is smaller than in the $v'-u'$ path.)
This is a contradiction to the assumption that $u' \in \Stagsqrt(v')$. (As there are at least $k$ closer than $u'$ to $v'$ selected vertices.)

Hence $h \le k$. For every $i \in [0,h-1]$, the number of hops on $P^{\sqrk}(v',u')$ between $v'_i$ and $v'_{i+1}$ is, conditioned on $\cA$, 
 at most $\sqrt{n}$. (Otherwise there were another selected  vertex between them on the path.) Hence there are edges $(v'_i,v'_{i+1})$, for every $i \in [0,h-1]$, in $G'$ of weight $\omega'(v'_i,v'_{i+1})$ equal to the weight $\omega(P(v'_i,v'_{i+1}))$ of the subpath $P(v'_i,v'_{i+1})$ of $P^{\sqrk}(v',u')$ between $v'_i$ and $v'_{i+1}$. Hence there is a $v'-u'$ path $(v'=v'_0,v'_1,\ldots,v'_{h-1},v'_h=u')$, $h \le k$, in $G'$ of weight $d_G^{\sqrk}(v',u')$.
\QED

\begin{lemma}
\label{lm:dist_eql}
Conditioned on $\cA$, for every $v' \in V'$, and for every $u' \in \Stagsqrt(v')$, we have $d_{G'}^{(k)}(v',u') = d_G^{\sqrk}(v',u')$.
\end{lemma}
\proof
By Lemma \ref{lm:path}, $d_{G'}^{(k)}(v',u') \le d_G^{\sqrk}(v',u')$.
Suppose for contradiction that there is a strict inequality. But then we can translate the $k$-limited $v'-u'$ path in $G'$ of weight $d_{G'}^{(k)}(v',u')$ into a $\sqrk$-limited $v'-u'$ path in $G$ of the same weight. This however results in a $\sqrk$-limited path between them of length strictly smaller than $d_G^{\sqrk}(v',u')$, contradiction.
\QED

\begin{lemma}
\label{lm:subsets}
Conditioned on $\cA$, for every $v' \in V'$,
$$\Stagtagk(v') \subseteq \Stagkk(v')~.$$
\end{lemma}
\proof
Consider $u' \in \Stagtagk(v')$. There exists a $\sqrk$-limited path $\pi(v',u')$ in $G$ of length at most $d_{G'}^{(k)}(v',u')$. 
(This path is obtained by concatenating $\sqrt{n}$-limited paths from $G$ that correspond to edges of the $k$-limited $v'-u'$ path in $G'$.) 

If $u' \nin \Stagkk(v')$, it follows that there are some other $k$ vertices $u'_1,u'_2,\ldots,u'_k \in \Stagkk(v')$, such that for all $i\in [k]$, $u'_i \neq u'$, and {\em either} 
$d_G^{\sqrk}(v',u'_i) < d_G^{\sqrk}(v',u')$ {\em or} $d_G^{\sqrk}(v',u'_i) = d_G^{\sqrk}(v',u')$ and the number of hops in the shortest $\sqrk$-limited $v'-u'_i$ path $\pi(v',u'_i)$ in $G$ is smaller than in the shortest $\sqrk$-limited $v'-u'$ path $\pi(v',u')$ in $G$, {\em or} also 
$|\pi(v',u'_i)| = |\pi(v',u')|$, but $\pi(v',u'_i) <_{v'} \pi(v',u')$.

Also, by Lemmas \ref{lm:path} and \ref{lm:dist_eql}, $d_G^{\sqrk}(v',u') = d_{G'}^{(k)}(v',u')$, and $d_G^{\sqrk}(v',u'_i) = d_{G'}^{(k)}(v',u'_i)$, for all $i \in [k]$. 
Hence,  for all $i \in [k]$, either $d_{G'}^{(k)}(v',u'_i) < d_{G'}^{(k)}(v',u')$,  or  $d_{G'}^{(k)}(v',u'_i) = d_{G'}^{(k)}(v',u')$, but
$|\pi(v',u'_i)| < |\pi(v',u')|$, or also $|\pi(v',u'_i)| = |\pi(v',u')|$, but $\pi(v',u'_i) <_{v'} \pi(v',u')$.
In either case, this is a contradiction to the assumption that $u' \in \Stagtagk(v')$.
(Recall that the ties in $\Stagtagk(v')$ are broken using respective paths in $G$, and not in $G'$.)
\QED

\begin{lemma}
\label{lm:subsets_opp}
Conditioned on $\cA$, for every $v' \in V'$,
$$\Stagkk(v')   \subseteq  \Stagtagk(v')~.$$
\end{lemma}
\proof
If $|\Stagtagk(v')| = k$, then since by Lemma \ref{lm:subsets}, $\Stagtagk(v') \subseteq \Stagkk(v')$, and 
$|\Stagkk(v')| \le k$, it follows that the two sets are equal. 

Otherwise, $|\Stagtagk(v')| < k$. Suppose for contradiction that there exists $u' \in \Stagkk(v') \setmns \Stagtagk(v')$.
By Lemma \ref{lm:dist_eql}, $d_{G'}^{(k)}(v',u') = d_G^{\sqrk}(v',u')$. Since $u' \in \Stagkk(v')$, by definition of this set, $d_{G'}^{(k)}(v',u') = d_G^{\sqrk}(v',u') < \infty$. But then $u'$ must have been included in $\Stagtagk(v')$, because $|\Stagtagk(v')| < k$ implies that for all vertices $u' \in V' \setmns \Stagtagk(v')$, we have $d_{G'}^{(k)}(v',u') = \infty$. This is a contradiction. 
\QED

\begin{corollary}
\label{cor:sets_eql}
Conditioned on $\cA$, for every vertex $v' \in V'$, 
$$\Stagkk(v')   = \Stagtagk(v')~.$$
\end{corollary}

Finally, we argue that for any vertex $v' \in V'$, $\Stagtagk(v') = \Stagtag(v')$. (The  latter set is the set of $k$ closest vertices in $G'$ to $v'$, where the paths are no longer $k$-limited. The ties are broken in the same way as before, i.e., using the respective paths in $G$.)
The proof of the following lemma is in Appendix \ref{app:pf}.

\begin{lemma}
\label{lm:Stagtag}
Conditioned on $\cA$, for every $v' \in V'$,
$$\Stagtagk(v') = \Stagtag(v')~.$$
\end{lemma}
\def\Pfstag
{
Consider a vertex $u' \in \Stagtag(v')$. Let $\pi_G(u',v')$ be the shortest $u'-v'$ path  in $G$ with the smallest number of hops.
Let $\pi'(u',v') = (u' = u'_0,u'_1,\ldots,u'_h = v')$ be the vertices of $V' \cap \pi_G(u',v')$, in the order of their appearrence on $\pi_G(u',v')$.
By a previous argument, for every $i \in [0,h-1]$, $(u'_i,u'_{i+1}) \in E'$. (This can be seen by repeatedly applying Lemma \ref{lm:sqrt} to $\pi_G(u',v')$. First, we consider a subpath $(u' = u^{(0)},u^{(1)},\ldots,u^{(\lceil \sqrt{n} \rceil)})$ of the first $\lceil \sqrt{n} \rceil$ edges from $G$ of $\pi_G(u,v)$. By Lemma \ref{lm:sqrt}, this subpath contains an internal vertex $u'_1 = u^{(j_1)} \in V'$, for some $j_1 \le \sqrt{n}$. Then we consider a $\lceil \sqrt{n} \rceil$-long subpath of $\pi_G(u,v)$ that starts at $u'_1$. By applying to it Lemma \ref{lm:sqrt} again, we obtain $u'_2$, etc.)

Observe that $h-1 < k$, as there are $h-1$ vertices $u'_1,\ldots,u'_{h-1} \in V'$ which are all closer to $v'$ than $u'$ in $G'$. (Or, if zero weights are allowed, then some of the $u'_i$ may be at the same distance from $v'$ as $u'$, but the number of hops in the shortest  $v'-u'$ path 
 in $G$ with the smallest number of hops is larger than in the respective path between $v'$ and $u'_i$.) Hence if $h-1 \ge k$, then $u' \nin \Stagtag(v')$, contradiction. 
 
Hence $u'$ is reachable from $v'$ in $G'$ within $h \le k$ hops, and moreover, $d_{G'}^{(k)}(u',v') = d_{G'}(u',v') < \infty$.
(Recall that $u' \in \Stagtag(v')$, and the latter set contains only vertices reachable from $v'$ in $G'$.)

If $u' \nin \Stagtagk(v')$, then since $d^{(k)}_{G'}(u',v') < \infty$, it follows that there exists vertices $u'_1,\ldots,u'_k \in V'$,
$u' \nin \{u'_1,\ldots,u'_k\} = \Stagtagk(v')$, such that for every $i \in [k]$, 
$$d_{G'}^{(k)}(v',u'_i) \le d_{G'}^{(k)}(v',u') = d_{G'}(v',u') < \infty~.$$
Also, $d_{G'}(v',u'_i) \le d_{G'}^{(k)}(v',u'_i)$, i.e., $d_{G'}(v',u'_i) \le d_{G'}(v',u')$, for every $i \in [k]$. Moreover, if there is an equality ($d_{G'}(v',u'_i) = d_{G'}(v',u')$), then $\pi_G(v',u'_i) <_{v'} \pi_G(u',v')$.

But this is a contradiction to the assumption that $u' \in \Stagtag(v')$. 
Hence 
\begin{equation}
\label{eq:one_dir}
\Stagtag(v') \subseteq \Stagtagk(v')~.
\end{equation}
Now we prove the opposite direction, i.e., $ \Stagtagk(v') \subseteq \Stagtag(v')$.

If $|\Stagtag(v')| = k$ then (\ref{eq:one_dir}) implies that the two sets are equal, because $|\Stagtagk(v')| \le k$.

So assume that $|\Stagtag(v')| < k$. Hence for every $u' \nin \Stagtag(v')$, we have $d_{G'}(v',u') = \infty$, and thus $d_{G'}^{(k)}(v',u') \ge d_{G'}(v',u') = \infty$ as well. Suppose for contradiction that there exists a vertex $u' \in \Stagtagk(v') \setmns \Stagtag(v')$.
Then $u' \nin \Stagtag(v')$ implies $d_{G'}^{(k)}(v',u') \ge d_{G'}(v',u')  = \infty$, and thus $u' \nin \Stagtagk(v')$ as well, contradiction.
(Recall that, by definition, the latter set contains only reachable vertices.)
\QED
}

Hence, conditioned on $\cA$, $\Stagkk(v') = \Stagtag(v')$, i.e., as a result of our computation, after $O(\sqrt{n} \cdot k^2)$ rounds every virtual vertex $v' \in V'$ knows its $\Stagtag(v')$. We also want to ensure that every vertex $u' \in V'$ that belongs to $\Stagtag(v')$, for some $v' \in V'$, knows about this hopset edge $(v',u')$. (We just showed that $v'$ does know about it.) 

To ensure this, we conduct an upcast and pipelined broadcast of all the computed edges $\{(v',u') \mid u' \in \Stagtag(v'), v' \in V'\} = G'^{(k)}$ over an auxiliary BFS tree $\tau$ of the entire graph $G$.
Since there are $O(|V'| \cdot k) = O(\sqrt{n} \cdot \log n \cdot k)$ edges in $G'^{(k)}$, it follows that this step requires $O(D + \sqrt{n} \cdot k \cdot \log n)$ time.

\begin{corollary}
\label{cor:hopset}
Whp, in overall time $O(D + \sqrt{n}\cdot k^2 + \sqrt{n}k \cdot \log n)$, the $k$-shortcut hopset ${G'}^{(k)}$ of the virtual graph $G'$ can be computed from scratch.
\end{corollary}

We remark that the virtual graph $G'$ itself is {\em not} known to the vertices of $V'$, even after the computation of the hopset $G'^{(k)}$ for $G'$ has been completed.

\section{Computing Distances}
\label{sec:dist}

In this section we employ the $k$-shortcut hopset computed in Section \ref{sec:hopset} to compute distances from a designated root vertex $r$ to all other vertices of the graph. We will also extend this algorithm to compute a shortest paths tree (henceforth, SPT) rooted at $r$.

We can assume that the root $r$ belongs to $V'$. (It can be just added to $V'$ after sampling all other vertices of $V'$. This has no effect on the analysis, except that the expected size of $V'$ grows by an additive 1.)
Our first objective at this stage is to compute all distances $\{d_G(r,v') \mid v' \in V'\}$. Specifically, we want every $v' \in V'$ to know its respective distance $d_G(r,v')$. The algorithm will utilize an auxiliary BFS tree $\tau$ of $G$ rooted at a vertex $\rt$. It can be constructed in $O(D)$ time. (See, e.g., \cite{Pel00:ln}, Chapter 5.) 

Recall that the hopbound of $\Gtagk$ is $h' = O(N/k) = O\left({{\sqrt{n} \log n} \over k}\right)$, whp.
This stage runs for $h'$ iterations, i.e., it is a B-F to depth $h'$ rooted at $r$, in $G' \cup \Gtagk$.
(By ``B-F to depth $h$'' we mean that the B-F explores vertices that are at {\em hop}-distance at most $h$ from the origin.)
 At the beginning of 0th iteration, 
every vertex $v' \in V' \setmns \{r\}$ initializes its estimate $\delta(v') = \infty$, and $r$ initializes its estimate $\delta(r) = 0$.
In iteration $i$, $0 \le i \le h'-1$, every virtual vertex $v' \in V'$ that has a finite estimate $\delta(v')$ wishes to deliver its estimate to all its neighbors in $G' \cup \Gtagk$.  
Before iteration $i$ starts, every vertex $v$ initializes two auxiliary  estimates $\delta^{I}(v) = \delta^{II}(v) = \delta(v)$.
The iteration lasts for $O(D + |V'|) + O(\sqrt{n}) = O(D + \sqrt{n} \log n)$ rounds, and it consists of two parts.
In the first part, all vertices $v'$ upcast their estimates to the root $\rt$ of the auxiliary tree $\tau$, and the root broadcasts them to the entire graph.
Every vertex $u' \in V'$ that hears an estimate $\delta(v')$ of its neighbor $v'$ in the hopset $\Gtagk$, computes its own estimate $\delta_{v'}(u') = \delta(v') + \omega^{(k)}(v',u')$, compares it with the minimum estimate $\delta^I(v')$ it knows, and updates the minimum estimate if needed. (Note, however, that on the second part of the iteration $v'$ still disseminates its original estimate $\delta(v')$ and not the updated estimate $\delta^I(v')$.)  This part of the computation requires $O(|D| + |V'|) = O(D + \sqrt{n} \log n)$ time. 

In the second part of the iteration, every vertex $v'$ initiates a B-F in $G$ to depth $\sqrt{n}$. The message $v'$ broadcasts is $\delta(v')$.
Every intermediate vertex $v \in V$ that hears from each of its neighbors $\{u_1,\ldots,u_d\}$ estimates $\{\delta(u_1),\ldots,\delta(u_d)\}$ of their respective distances from $r$, computes the smallest value $\delta(u_i) + \omega((u_i,v))$, compares it with its current estimate $\delta(v)$, sets the minimum as its new estimate, and sends the latter to its neighbors. (We also want every vertex to record its {\em parent}, i.e., the vertex from which it learnt its current estimate.) This process continues for $\sqrt{n}$ rounds. In each round, every vertex sends one message to all its neighbors. 

Since each edge $e = (v',u')$ of $G'$ corresponds to a $\sqrt{n}$-limited $v'-u'$ path in $G$, this $\sqrt{n}$-limited B-F exploration serves to imitate one phase of a B-F exploration in $G'$. Specifically, as a result of the second part of iteration $i$, every vertex $v' \in V'$ learns the value 
$$\delta^{II}(v') = \min\{\delta(u') + d_G(v',u') \mid (v',u') \in E'\} = \min\{\delta(u') + \omega'(v',u') \mid (v',u') \in E'\}~.$$
Now $v'$ computes the minimum between $\delta^I(v')$ and $\delta^{II}(v')$, and sets it as its new estimate $\delta(v')$. The latter estimate will be used in the next iteration of the B-F in $G \cup \Gtagk$.

Observe that in the second part of each iteration, we essentially run the B-F in $G$ that we used to compute the hopset $\Gtagk$, but with $k=1$. (But we deliver just one single smallest value, and not the $k$ smallest ones. Also, now every vertex $v$ is interested in its distance from a designated root $r$, rather than in the distances from $k$ closest vertices of $V'$.)

Hence a special case of the analysis that we used in Section \ref{sec:hopset} shows that after $i$, $0 \le i \le \sqrt{n}$, rounds of this B-F, every vertex $v \in V$ knows $\delta^{(i)}(v) = \min \{\delta(v') + d_G^{(i)}(v',v)\}$, where the minimum is taken over all $v' \in V'$ reachable from $v$ via at most $i$ hops in $G$.  In particular, 
 after $\sqrt{n}$ rounds of this process, every $u' \in V'$ knows its $\delta^{(\sqrt{n})}(u')$.

Hence, overall, iteration $i$ requires $O(D + \sqrt{n} \log n)$ time, and it imitates one iteration of B-F in $G' \cup \Gtagk$, rooted at $r$. Since the hopbound of $\Gtagk$ is $h' = O(N/k) = O\left({{\sqrt{n} \log n} \over k}\right)$, it follows that within $h'$ such iterations, every vertex $v' \in V'$ knows its correct distance estimate $d^{(h')}_{G' \cup \Gtagk}(r,v') = d_{G'}(r,v') = d_G(r,v')$.
(The last equality holds whp. It is true because, by Lemma \ref{lm:sqrt}, the shortest $r-v'$ path in $G'$ contains vertices of $V'$ every $\sqrt{n}$ hops or less, i.e., it can be replaced by a $r-v'$ path in $G'$ of the same length.)

\begin{corollary}
\label{cor:Vtag_dists}
After the hopset $\Gtagk$ has already been constructed, within additional
$O(D + \sqrt{n} \log n) \cdot O\left({{\sqrt{n} \log n} \over k}\right)$ time, all distances $\{d_G(r,v') \mid v' \in V'\}$ can be computed.
\end{corollary}

Hence the overall time spent by the algorithm so far (including the time required to construct the $k$-shortcut hopset $\Gtagk$) is 
$O(D + \sqrt{n} \log n)\cdot  O\left({{\sqrt{n} \log n} \over k}\right) + O(\sqrt{n} \cdot k^2 + \sqrt{n} \cdot  k \log n)$.

Now, when all virtual vertices $v' \in V'$ know their exact disatnces $d_G(r,v')$ from $r$, we conduct a $\sqrt{n}$-limited B-F exploration in $G$ from all vertices of $V'$. (Every vertex $v \in V$, in every round, selects one single smallest estimate of its distance from $r$, and forwards it.) This step requires $\sqrt{n}$ additional time.

\begin{lemma}
\label{lm:all_dists}
Conditioned on $\cA$ (i.e., whp), for every vertex $v \in V$, after the last step of the algorithm we have $\delta(v) = d_G(r,v)$. 
\end{lemma}
\proof
Let $\pi(r,v)$ be a shortest $r-v$ path in $G$, with the smallest number of hops, smallest with respect to $<_v$ among such paths. 

First, consider the case that $|\pi(r,v)| \le \sqrt{n}$. Recall that $r \in V'$. Then the distance $d_G(r,v)$ propagates from the root $r$ to $v$ along $\pi(r,v)$ during the last stage of the algorithm (i.e., during the $\sqrt{n}$-limited B-F), and we are done.

Now we turn to the case $|\pi(r,v)| > \sqrt{n}$.       Let $(r = v'_0,v'_1,\ldots,v'_h,v)$, $v'_0,v'_1,\ldots,v'_h \in V'$ be the virtual vertices
(i.e., vertices of $V'$) appearing on $\pi(r,v)$, in the order of their appearance. Under $\cA$, for every index $i \in [0,h-1]$, the number of hops between $v'_i$ and $v'_{i+1}$ in $\pi(r,v)$ is at most $\sqrt{n}$, and so is the number of hops between $v'_h$ and $v$. At the last stage of the algorithm, the vertex $v'_h$ holds $\delta(v'_h) = d_G(r,v'_h)$. 

Denote by $\pi(v'_h,v)$ the subpath of $\pi(r,v)$ connecting $v'_h$ and $v$. The estimate $\delta(v'_h)$ propagates along $\pi(v'_h,v)$ during the $\sqrt{n}$-limited B-F on the last stage of the algorithm, and at the end of the algorithm it holds that
$$
\delta(v) ~\le~ \delta(v'_h) + \omega(\pi(v'_h,v)) ~=~ d_G(r,v'_h) + d_G(v'_h,v) ~=~d_G(r,v)~.
$$
The last inequality is because $v'_h$ lies on the shortest $r -v$ path in $G$.

Suppose for  contradiction  that $\delta(v) < d_G(r,v)$. But then there exists some virtual vertex $v' \in V'$, such that $\delta(v) = \delta(v') + d_G(v',v)$. (This is the virtual vertex through which $v$ has acquired its estimate $\delta(v)$.) 
But $\delta(v') = d_G(r,v')$, and so there exists an $r -v$ path in $G$ that passes through $v'$ and has length 
$\delta(v) = d_G(r,v') + d_G(v',v) < d_G(r,v)$, contradiction.
\QED

Next we show that the algorithm can also construct an SPT for $G$ rooted at $r$. In the beginning  we assume, for convenience of presentation, that all edge weights are positive.

The modification to the algorithm is in the last stage, where a $\sqrt{n}$-limited B-F in $G$ from vertices of $V'$ is conducted. 
There are two modifications. First, every vertex $v \in V$ records the neighbor $p(v)$ of $v$ in $G$ from which it received its final distance estimate.  (For this end it always keeps the neighbor that supplied $v$ its current estimate; at the end this neighbor is $p(v)$.)
Second, every vertex $v' \in V'$ will now also receive updates on this last stage of the algorithm. (This is not necessary if one is only in interested in distances.) The vertex $v'$ will record the neighbor $p(v')$ of $v'$ in $G$, through which $v'$ could receive a correct estimate $\delta(v') = d_G(r,v)$. 
That is, every neighbor $u$ of $v'$ sends it some value $\delta(u)$, and $v'$ computes $\min\{\delta(u) + \omega((u,v')) \mid u \in \Gamma_G(v')\}$. This value is equal to the value $\delta(v') = d_G(r,v')$, which $v'$ knows before the last phase begins. Nevertheless, the vertex $v'$ records the neighbor $p(v') = u$ through which it can attain this value. 
Ties are broken in the same way as above, i.e., according to the number of hops between $r$ and $p(v)$, and finally, in case of equality, using the $\ID$ numbers of neighbors $p(v)$. The argument that shows that this is indeed the same value is given in Lemma \ref{lm:all_dists}.

We now argue that the resulting edge set is an SPT of $G$ with respect to $r$.

\begin{lemma}
\label{lm:spt}
The edge set $T =\{(v,p(v)) \mid v \in V \setmns \{r\}\}$ is an SPT of $G$ with respect to $r$.
\end{lemma}
\proof
The edge set $T$ spans $V$. Moreover, it is acyclic, as $\delta(p(v)) < \delta(v)$, for all $v \in V \setmns \{r\}$. (Because edge weights are positive.)
This also guarantees that if $u = p(v)$ then it cannot happen that $p(u) = v$, and so the number of edges in $T = \{(v,p(v)) \mid v \neq r\}$ is $n-1$. 
 Hence $T$ is a spanning tree of $G$. 

The proof that $T$ is an SPT is by induction on the depth (the hop-distance from $r$) in $T$ of a vertex $v$.
The base case  ($v = r$) clearly holds. 

For the induction step, observe that
by the induction hypothesis,
$
d_T(r,p(v)) = d_G(r,p(v))$. Also, 
$
d_T(r,v) = d_T(r,p(v)) + \omega((p(v),v))  = d_G(r,p(v)) + \omega((p(v),v))$.
But the latter is equal to the distance estimate $\delta(v) = \delta(p(v)) + \omega((p(v),v))$ of $v$ (from $r$), and we have already established that $\delta(v) = d_G(r,v)$. Hence $d_T(r,v) = d_G(r,v)$.
\QED

This argument can be extended to the case that zero weights are allowed, in the following way. Let $T'$ be an SPT of $G' \cup \Gtagk$ with respect to $r$, that our algorithm implicitly constructs when it computes the distances $\{d_G(r,v') \mid v' \in V'\}$. Every vertex $v' \in V'$ will record its hop-distance $h'(v') \le h'$ from $r$ in $T'$, i.e., the number of phase of the B-F over $G' \cup \Gtagk$ in which it acquired its final distance estimate. Also, in the last stage of the algorithm, a vertex $v$ will prefer as its parent $p(v)$ the vertex $u$ with smallest $\delta(u) + \omega((u,v))$, and among such vertices $v$ will prefer $u$ which acquired its estimate through a virtual vertex $v'$ with smallest $h'(v')$.
Also, breaking ties among two such $v'_1,v'_2$ with equal $h'(v'_1) = h'(v'_2)$, the vertex $v$ will prefer the one with smaller number of hops  in $G$
between $v$ and this virtual vertex. (This corresponds to the iteration number of the last stage of the algorithm.) Finally, if those are equal, one breaks ties by $\ID$s of $v'_1,v'_2$, and for two messages that come from the same $v'$ via different neighbors $u_1,u_2$, the identities of $u_1,u_2$ will be used to break ties. 

With this choice of parents, it is easy to see that the resulting $T$ is acyclic, and contains $n-1$ edges. The rest of the proof of Lemma \ref{lm:spt} extends seamlessly.

The overall running time of the algorithm is, by Corollaries \ref{cor:hopset} and \ref{cor:Vtag_dists}, 
\begin{eqnarray*}
&&O(\sqrt{n} \cdot k^2 + D + \sqrt{n} \cdot k \cdot \log n) + (O(D + \sqrt{n} \cdot \log n) + O(\sqrt{n}))\cdot O({{\sqrt{n} \log n} \over k}) ~= \\
&& O(D \cdot {{\sqrt{n} \cdot \log n} \over k} + \sqrt{n} \cdot k^2 + {{n \cdot \log^2 n} \over k} + \sqrt{n} \cdot k \cdot \log n)~.
\end{eqnarray*}
Set $k = n^{1/6} \cdot \log^{2/3} n$. We obtain time $O((D + n^{1/2} \cdot \log n) \cdot (n \cdot \log n)^{1/3})$. This is $O(n^{5/6} \cdot \log^{4/3} n)$, for $D = O(n^{1/2} \cdot \log n)$. This bound is sublinear in $n$ for $D = o(n^{2/3}/\log^{1/3} n)$. 

However, when $D$ is large, i.e., $D = \Omega(\sqrt{n} \cdot \log n)$, it makes sense to set $k = (D  \cdot \log n)^{1/3}$. Observe that $k = (D \cdot \log n)^{1/3} \le \sqrt{n} \cdot \log n = N$, and so this is a valid choice of the parameter $k$. 
With this choice of $k$, the running time becomes 
$$O(\sqrt{n} \cdot D^{2/3} \cdot \log^{2/3} n) +  O({ {n \cdot \log^{5/3} n} \over {D^{1/3}}}) ~=~ O(D^{2/3} \sqrt{n} \cdot \log^{2/3} n)~.$$
(The last inequality is because $D = \Omega(\sqrt{n} \cdot \log n)$.)
This estimate is sublinear in $n$ as long as $D = o({{n^{3/4} } \over {\log n}})$. It is also no worse  than the estimate $O((D + \sqrt{n} \log n) \cdot (n \log n)^{1/3})$ in the entire range $D = \Omega(\sqrt{n} \log n)$.

To summarize:

\begin{theorem}
\label{thm:basic}
Whp, the algorithm described above computes an exact single source shortest paths tree in the CONGEST model in $O(n^{5/6} \cdot \log^{4/3} n)$ time, whenever $D = O(\sqrt{n} \cdot \log n)$, and in $O(D^{2/3} \sqrt{n} \cdot \log^{2/3} n)$ time, for larger $D$.
\end{theorem}

See Theorem \ref{thm:single_src} for yet sharper bounds.

\section{Improved Bounds}
\label{sec:impr}

In this section we show that by setting the sampling probability $q$ more carefully, one obtains yet sharper bounds.  Most notably, in this way we derive a variant of our algorithm that runs in sublinear in $n$ time, for a much wider range of $D$. Specifically, the running time will be sublinear in $n$ for $D = o(n/\log^2 n)$. 

The expected number of selected vertices $\Expect(|V'|) = n \cdot q$, i.e., whp, $|V'| = O(n \cdot q)$.
(We assume that $q = \Omega({{\log n} \over n})$.) Consider a path of ${n \over {n \cdot q}} \cdot (c \cdot \ln n) = {{c \ln n} \over q}$ hops, for a sufficiently large constant $c$.
It contains no selected vertex with probability $(1 -q)^{{c \ln n} \over q} \le {1 \over {n^c}}$. So, whp, (i.e., with probability at least $1 - {1 \over {n^{c-3}}}$), for every pair $u,v \in V$ of  vertices and an index $i$ such that $P^{(i)}(u,v)| \ge {{c\ln n} \over q}$,   the path $P^{(i)}(u,v)$ contains a selected vertex.

We conduct a B-F from all vertices of $V'$ to depth ${{c \cdot \ln n} \over q} \cdot k$, while storing and forwarding $k$ smallest values at every super-round. The total time required for this step is $O({{\ln n} \over q} \cdot k^2)$.

The upcast and pipelined broadcast (over the BFS tree $\tau$ of $G$) of the hopset edges requires $O(D + n \cdot q \cdot k)$ time.

We then conduct B-F in $G' \cup \Gtagk$ for $h' = O(N/k) = O({{n \cdot q} \over k})$ phases.
\begin{lemma}
This step requires $O(((D + n q) + {{\ln n} \over q}) \cdot {{nq } \over k})$ time.
\end{lemma}
\proof
Recall that every phase of this B-F consists of two parts. In the first part, $O(nq)$ estimates of distances of selected vertices from $r$ are disseminated over the BFS tree $\tau$. This requires $O(D + nq)$ time.
In the second part, a B-F exploration over $G$ to depth ${{c \ln n} \over q}$ is conducted from vertices of $V'$, where each vertex forwards just the smallest estimate of distance from $r$ that it knows. This requires $O({{\ln n} \over q})$ time.
 Since there are $O({{n \cdot q} \over k})$ phases of the B-F in $G' \cup \Gtagk$, it follows that the overall time of this step is  $O(((D + n q) + {{\ln n} \over q}) \cdot {{nq } \over k})$.
\QED

Finally, when all virtual vertices (of $V'$) already know their respective distances from $r$, another B-F in $G$ to depth ${{c \ln n} \over q}$ is conducted, to update all vertices of $V \setmns V'$. This step requires  additional $O({{\ln n} \over q})$ time.

Hence the total running time of the algorithm is given by
\begin{equation}
\label{eq:time}
T ~=~ O\left({{\ln n} \over q} \cdot k^2 + (D + nq \cdot k) + (D + nq + {{\ln n} \over q}) \cdot {{nq} \over k}\right)~.
\end{equation}
\begin{comment}
The message complexity is given by 
\begin{equation}
\label{eq:msgs}
C ~= ~ O\left(|E| \cdot{ {\ln n} \over q} \cdot k^2 + n^2 q k + (n^2 q + |E| {{\ln n} \over q}){{nq} \over k}\right)~. 
\end{equation}
\end{comment}
To optimize the running time, we consider two regimes. The first regime is when $D$ is relatively small, i.e., $D = O(\max\{nq,{{\ln n} \over q}\})$.
We then substitute $q = \sqrt{{{\ln n}  \over n}}$, $k = (n \ln n)^{1/6}$. (Observe that $k \le nq$. This is required because $N = \Theta(n q)$, and $k$ in the $k$-shortcut hopset needs to be at most $N-1$.)
The bound becomes  $T = O((n \ln n)^{5/6})$  (for $D = O(\sqrt{n \ln n})$).
This slightly improves the bound $T = O(n^{5/6} \cdot \log^{4/3} n)$ that we had in Theorem \ref{thm:basic}. 

Next we consider the regime of large $D$, i.e., $D \ge \max\{{{\ln n} \over q}, nq\}$. 
We set $q = {{\ln n} \over D}$, $k = \left({{n \ln n} \over D}\right)^{1/3}$. (Note that $nq \ge k$, as required.)  We also have $D = {{\ln n} \over q} \ge nq$ in this case, as $D= \Omega(\sqrt{n \log n})$. 
Then the running time is bounded by 
\begin{eqnarray*}
T &=& O\left({{\ln n} \over q} \cdot k^2 + (D + nq \cdot k) + D \cdot {{nq } \over k}\right) ~=~ O\left(D^{1/3} (n \ln n)^{2/3} + 
\left({{n \ln n} \over D}\right)^{4/3}\right) \\
&= & O( D^{1/3} (n \ln n)^{2/3})~.
\end{eqnarray*}
(The last inequality is because $D = \Omega(\sqrt{n \log n})$.)

Note that for $D = \Theta(\sqrt{n \log n})$, this estimate gives $T = O((n \log n)^{5/6})$, i.e., it agrees with the bound that we have in the small diameter regime. Also, this bound is sublinear in $n$ as long as $D = o(n/\log^2 n)$. 
\begin{comment}
The message complexity of the algorithm behaves as
$$C ~=~ O(|E| \cdot D^{1/3} (n\ln n)^{2/3} + {{n^{7/3} (\ln n)^{4/3}} \over {D^{4/3}}}) ~=~ O(|E| \cdot D^{1/3} (n\ln n)^{2/3})~.$$
(The last inequality is because $D = \Omega(\sqrt{n \log n})$.) Hence the message complexity of our algorithm is $o(|E| n)$, under the same (mild) constraint $D = o(n/\log^2 n)$ on the diameter. Hence in this range of parameters ($D = o(n/\log^2 n)$), our algorithm strictly out-performs the  Bellman-Ford algorithm both in terms of time and message complexities.
\end{comment}
  
We summarize this analysis with the next theorem.

\begin{theorem}
\label{thm:single_src}
Whp, our algorithm computes an exact shortest paths tree in the $\CONGEST$ model in time $O((n \log n)^{5/6})$,
when $D = O(\sqrt{n \log n})$, and in time $O(D^{1/3} (n \log n)^{2/3})$, for larger $D$.
\end{theorem}

\section{Multiple Sources}
\label{sec:mult_src}

In this section we extend our algorithm so that it will compute  shortest paths between  $s$ sources $r_1,\ldots,r_s$ and all other vertices.
The algorithm still builds the $k$-shortcut hopset $\Gtagk$ in the same way as in the single-source case. The hopbound  is still $h' =O(N/k) = O(nq/k)$. The time required to construct it is, as was shown above, 
$O({{\log n} \over q} \cdot k^2) + O(D + n q k)$.

We will assume that $s \le nq$, and so the sources $\{r_1,\ldots,r_s\}$ can be added to $V'$ without affecting its size by more than a constant factor. 

Next, we conduct a B-F in $G' \cup \Gtagk$ from the $s$ sources $r_1,\ldots,r_s$. As before, this B-F continues for $h' = O(nq/k)$ iterations, and each iteration consists of two parts. In the first part all vertices $v' \in V'$ upcast their $s$ distance estimates via the BFS tree $\tau$ of $G$ to the root of $\tau$, and then these estimates are disseminated in the graph via pipelined broadcast in $\tau$. This part requires $O(D + N \cdot s) = O(D + n\cdot q \cdot s)$ time.

In the second part of each iteration, a B-F in $G$ to depth $O\left({{\log n} \over q}\right)$ is conducted. Unlike the single-source variant of the algorithm, here every step of this B-F is a super-round that consists of $s$ rounds. Every vertex $v$ uses these $s$ rounds to update its neighbors with the at most $s$ estimates of its distances from the $s$ sources. Hence this B-F requires $O\left({{\log n} \over q} \cdot s\right)$ time. 

Thus, a single iteration of the B-F in $G' \cup \Gtagk$ requires $O(D + (nq + {{\log n} \over q}) \cdot s)$ time, and overall this B-F requires
$$O(D + (nq + {{\log n} \over q}) \cdot s) \cdot h' ~=~ O((D + (nq  + {{\log n} \over q}) \cdot s) \cdot {{nq} \over k})$$
time.

Finally, on the last stage of the algorithm, vertices of $V \setmns V'$ learn their distances to the $s$ sources via a B-F in $G$ to depth $O\left({{\log n} \over q}\right)$. Each step of this B-F requires now $O(s)$ rounds, as $s$ estimates need to be proliferated. Hence the running time  of this step is $O\left({{\log n} \over q} \cdot s\right)$.

To summarize, the running time of the entire algorithm becomes 
\begin{equation}
\label{eq:T_multiple}
T = O\left({{\log n} \over q} \cdot k^2 + D + nq k + (D + (nq + {{\log n} \over q}) \cdot s) \cdot {{nq } \over k} + {{\log n} \over q} \cdot s\right)~.
\end{equation}
To analyze this expression, we again consider two regimes. The first regime is when $D$ is relatively small, i.e., $D \le s(nq + {{\log n} \over q})$. Here we set $q = \sqrt{{\log n} \over n}$, $k = (n \log n)^{1/6} \cdot s^{1/3}$. 
The condition $k \le nq$ holds. The running time becomes
\begin{eqnarray*}
T &=& O((n \log n)^{5/6} \cdot s^{2/3} + (n \log n)^{2/3} \cdot s^{1/3} + s \cdot \sqrt{n \log n} \cdot {{(n \log n)^{1/3} } \over {s^{1/3}}} + \sqrt{n \log n} \cdot s) \\
& = & O((n \log n)^{5/6} \cdot s^{2/3})~.
\end{eqnarray*}
This bound is applicable as long as $D = O(s \sqrt{n \log n})$, $s \le nq = O(\sqrt{n \log n})$. (We will handle the case of larger $s$ below.)
Note that this bound is better than the only previously existing trivial bound $O(n \cdot s)$ in the entire range where it applies.

The second regime is when $D = \Omega(s (nq + {{\log n} \over q}))$, i.e., in particular, $D = \Omega(s \sqrt{n \log n})$. By (\ref{eq:T_multiple}), 
the running time in this regime is given by $T = O\left({{\log n} \over q} \cdot k^2 + nq k + D \cdot {{nq} \over k}\right)$. 
We set $q = {{s \log n} \over D} \le 1$ (as $D \ge s \log n$), and $k = \left({{n \log n} \over D}\right)^{1/3} \cdot s^{2/3}$. Note that $k \le nq$ holds. We conclude that
\begin{eqnarray*}
T & = & O\left({{\log n} \over {s \log n}} \cdot D \cdot \left({{n \log n} \over D}\right)^{2/3} \cdot s^{4/3}  + {{n s \log n} \over D} \cdot 
\left({{n \log n} \over D}\right)^{1/3} \cdot s^{2/3} + D \cdot {{n s \log n \cdot D^{1/3}} \over {D \cdot (n \log n)^{1/3} \cdot s^{2/3}}}\right) \\
& = & O((Ds) ^{1/3} (n \log n)^{2/3}  + \left({{n \log n} \over D}\right)^{4/3} \cdot s^{5/3})~.
\end{eqnarray*}
The second term is dominated by the first one when 
\begin{equation}
\label{eq:cond}
D ~\ge ~ s^{4/5} (n \log n)^{2/5}~.
\end{equation}
Since we are in the regime that $D \ge s \sqrt{n \log n}$, the condition (\ref{eq:cond}) holds. Hence $T = O((Ds)^{1/3} \cdot (n \log n)^{2/3})$.

We summarize this result in the next theorem. (This is the case of a relatively small $s$, i.e., $s = O(\sqrt{n \log n})$.)

\begin{theorem}
\label{thm:few_srcs}
Whp, our algorithm computes exact  shortest paths for pairs in $S \times V$, $|S| =s$, in the $\CONGEST$ model in time $O((n \log n)^{5/6} \cdot s^{2/3})$, whenever $D = O(s \sqrt{n \log n})$, $s = O(\sqrt{n \log n})$, and in time $O((Ds)^{1/3} (n \log n)^{2/3})$, for $D  = \Omega(s \sqrt{n \log n})$. (In the latter case, in particular, that $s = O(\sqrt{n/\log n})$.)
\end{theorem}

Note that the two bounds agree when $D = \Theta(s \sqrt{n \log n})$. Also,  this theorem generalizes Theorem \ref{thm:single_src}. 
%

In the case $s \ge \sqrt{n \log n}$, we set $q = s/n$, $k = {{s^{4/3}} \over {(n \log n)^{1/3}}}$. 
Observe that $s \le nq = s$, and $k = {{s^{4/3}} \over {(n \log n)^{1/3}}} \le nq = s$, for all $1 \le s \le n$. 
It also holds that $nq = s \ge {{\log n} \over q} = {{\log n} \over s} \cdot n$, i.e., $s \ge \sqrt{n \log n}$.
Note also that $D = o\left( s(nq + {{\log n} \over q})\right)$ in this case, as the right-hand-side is $\omega(n)$.

Hence, by (\ref{eq:T_multiple}),  the running time is
\begin{equation}
\label{eq:larger_T}
T = O\left({{\log n} \over q} \cdot k^2 + nq k + s \cdot {{n^2 q^2} \over k}\right) ~=~ O((n \log n)^{1/3} \cdot s^{5/3})~.
\end{equation}
(Note that $nq k = {{s^{7/3}} \over {(n \log n)^{1/3}}} \le s^{5/3} (n \log n)^{1/3}$, for all $1 \le s \le n$.)

This bound is not trivial (i.e., $o(n s)$) almost in the entire range of parameters, specifically, for $s = o(n/\sqrt{\log n})$.
Observe that for $s = \Theta(\sqrt{n \log n})$, this bound agrees with the bound of Theorem \ref{thm:few_srcs}, and gives running time $O((n \log n)^{7/6})$. 
A better bound for large $s$ (i.e., $s = \Omega(\sqrt{n \log n})$) can be derived by partitioning the set $S$ into $\alpha = \lceil s/\sqrt{n\log n} \rceil$ subsets $S_1,S_2,\ldots,S_\alpha$ of sizes $O(\sqrt{n\log n})$ each, and  running this algorithm first for $S_1 \times V$, then for $S_2  \times V$, $\ldots$, and finally, for $S_\alpha \times V$. The overall running time of the algorithm becomes 
$O((n \log n)^{7/6} \cdot {s \over {{\sqrt{n \log n}}}}) = O((n \log n)^{2/3} \cdot s)$.
This bound is no worse than the bound (\ref{eq:larger_T}) in the entire range $s = \Omega(\sqrt{n \log n})$, and it is better than the bound in (\ref{eq:larger_T}) for $s = \omega(\sqrt{n \log n})$. We summarize this discussion below.

\begin{theorem}
\label{thm:many_srcs}
Whp, our algorithm computes exact  shortest paths for $S \times V$, $|S| = s$, in the $\CONGEST$ model in time $O( (n \log n)^{2/3} \cdot s)$, whenever $s = \Omega(\sqrt{n \log n})$, for all values of $D$. 
\end{theorem}

In particular, our algorithm computes {\em all-pairs shortest paths} in time $O(n^{5/3} \cdot \log^{2/3} n)$, for all values of $D$. 

Note also that the bound of Theorem \ref{thm:many_srcs} outperforms the second bound of Theorem \ref{thm:few_srcs}.
Indeed, $(Ds)^{1/3} \cdot (n \log n)^{2/3} \le s \cdot (n \log n)^{2/3}$ only if $D \le s^2$.
But the second bound of Theorem \ref{thm:few_srcs} applies only if $D = \Omega(s \sqrt{n \log n})$, i.e., $D = \Omega(\sqrt{D n \log n})$.
For the latter to hold, $D$ needs to be $\Omega(n \log n)$, i.e., this never happens.

\section{A Streaming Algorithm}
\label{sec:stream}

In this section we present a variant of our algorithm for computing exact shortest paths in the multi-pass streaming setting.

\subsection{Undirected Graphs}
\label{sec:stream_undir}

We start with undirected graphs, and then proceed to directed ones.

Fix an integer  parameter $k$, $1 \le k \le n-1$. Every vertex $v$ learns in one pass the $k$ closest neighbors $u_1,u_2,\ldots,u_k$ of $v$ in $G$, with ties broken by identity numbers.
(If $v$ happens to have degree smaller than $k$, then it learns all its neighbors.) This requires $O(n \cdot k)$ memory.
As a result we obtain the {\em $k$-neighborhood graph} $\cN = (V,F)$, $F = \{(v,u_i) \mid v \in V, i \in [k]\}$.

\begin{lemma}
\label{lm:path_contained}
For a vertex $v \in V$, and $u \in S_G[k](v)$, any shortest path $P(v,u)$ between $v$ and $u$ in $G$ with minimum number of hops is contained in the graph $\cN$.
\end{lemma}
\proof
Denote $P(v,u) = (v = v_0,v_1,\ldots,v_h = u)$, for some $h \le k$. 
(Note that if $h > k$, then $u \nin S_G[k](v)$, as the vertices $v_1,v_2,\ldots,v_k$ would have been preferred over $u$. This is a contradiction to the assumption that $u \in S_G[k](v)$.)
For some $i \in [0,h-1]$, we argue that $v_{i+1}$ is among $k$ closest neighbors of $v_i$, and thus the edge $(v_i,v_{i+1})$ belongs to the edge set $F$ of the $k$-neighborhood graph $\cN$. 

Indeed, otherwise there are neighbors $x_1,x_2,\ldots,x_k$ of $v_i$, which are all closer to $v_i$ than $v_{i+1}$. (Alternatively, they may be at the same distance from $v_i$, but have a smaller $\Id$ than that of $v_{i+1}$.) But then all these vertices are closer to $v$ than $v_{i+1}$ as well, and consequently, they are also closer to $v$ than $u = v_h$. (Note that $h \ge i+1$.) This is, however, a contradiction to the assumption that $u \in S_G[k](v)$. Hence $(v_i,v_{i+1}) \in F$. 
\QED

Note that for the sake of the definition of $S_\cN[k](v)$, the ties are broken in $\cN$ the same way as in $G$.

\begin{lemma}
\label{lm:same_sets}
For every $v \in V$, we have $S_\cN[k](v) = S_G[k](v)$.
\end{lemma}
\proof
First, we show that $S_G[k](v) \subseteq S_\cN[k](v)$.
Let $u \in S_G[k](v)$. By Lemma \ref{lm:path_contained}, any shortest $v-u$ path in $G$ with minimum number of hops is contained in $\cN$, and so, $d_\cN(v,u) = d_G(v,u) < \infty$. 
Also, recall that $S_G[k](v)$ contains only vertices $u$ reachable from $v$, and thus, such a path exists.

If $u \nin S_\cN[k](v)$, then there are $k$ vertices $\{y_1,\ldots,y_k\} = S_\cN[k](v)$, $u \nin \{y_1,\ldots,y_k\}$, with 
$d_\cN(v,y_i) \le d_\cN(v,u)$, and such that in case of equality, $y_i$ is preferred over $u$ in $\cN$.
But as $d_G(v,y_i) \le d_\cN(v,y_i)$, for every $i \in [k]$, we have 
$$d_G(v,y_i) ~\le~ d_\cN(v,y_i) ~\le~ d_\cN(v,u) ~=~d_G(v,u)~,$$
and in case of equality ($d_G(v,y_i) = d_G(v,u)$), the vertex $y_i$ is preferred over $u$ in $G$ too.
But then $u \nin S_G[k](v)$, contradiction. 

Next, we argue that $S_\cN[k](v) \subseteq S_G[k](v)$. 

If $|S_G[k](v)| = k$, then since $|S_\cN[k](v)| \le k$ and $S_G[k](v) \subseteq S_\cN[k](v)$, it follows that the two sets are equal.

Otherwise $|S_G[k](v)| < k$. Let $u \in S_\cN[k](v)$, and suppose for contradiction that $u \nin S_G[k](v)$. But $u$ is reachable from $v$ in $\cN$, and so it is reachable from $v$ in $G$ too, i.e., $d_G(v,u) < \infty$. 
On the other hand, when $|S_G[k](v)| < k$, it means that for every vertex $w \in V \setmns S_G[k](v)$, we have $d_G(v,w) = \infty$.
This is a contradiction.  Hence $S_\cN[k](v) \subseteq S_G[k](v)$, proving the lemma.
\QED

\begin{corollary}
\label{cor:stream_hopset}
In one pass over the stream, using $O(n k)$ memory, one can compute the $k$-shortcut hopset $G^{(k)}$.
\end{corollary}
\proof
We saw that in one pass, using $O(nk)$ memory, one computes the $k$-neighborhood graph $\cN$, and that for every $v \in V$, 
we have $S_G[k](v) = S_\cN[k](v)$. Given the graph $\cN$, we are now computing $\{S_\cN[k](v) \mid v \in V\}$ offline (i.e., without any additional passes over the stream), and obtain as a result the sets $S_G[k](v)$, for every vertex $v \in V$. 
Recall that the edge set of the hopset  $G^{(k)}$   is $\{(v,u) \mid u \in S_G[k](v), v \in V\}$. The algorithm has computed it.
\QED

Recall that $G^{(k)}$ is an exact hopset with hopbound $h= O(n/k)$. So now we conduct $h$ additional passes over $G$, and after each pass we relax also the edges of $G^{(k)}$. (In other words, we conduct Bellman-Ford in $G \cup G^{(k)}$ for $h = O(n/k)$ iterations.) Hence, after $i$ passes, $0 \le i \le h$, we have computed $i$-limited distances in $G \cup G^{(k)}$ from a designated root vertex $r$ to all other vertices. (For every vertex $v$, we store only its current distance estimate, and a parent through which this estimate was attained.) Hence, after $h = O(n/k)$ passes, we have computed $O(n/k)$-limited distances from the root $r$ in $G \cup G^{(k)}$, which are equal (since $G^{(k)}$ is an exact hopset with hopbound $O(n/k)$) to exact $\{r\} \times V$ distances in $G$. 

Observe that when computing hopset edges $\{(v,u) \mid v \in V, u \in S_\cN[k](v) = S_G[k](v)\}$, the algorithm has also computed the shortest paths $P(v,u)$ in $G$ implementing these hopset edges. We store them  implicitly, i.e., for every vertex $v$, we store a shortest paths tree (SPT) of $S_\cN[k](v) = S_G[k](v)$. (Note that its size is at most $k$.)
Once we have computed the SPT in $G \cup G^{(k)}$ rooted at $r$, containing at most $n-1$ edges, we can now replace every hopset edge of this tree with a path of length at most $k$, consisting of edges from $G$. In this way, we recover the shortest paths in $G$,  while still employing $O(n k)$ memory. 

\begin{theorem}
\label{thm:stream_spt}
For any $n$-vertex weighted undirected graph $G = (V,E)$, and any integer parameter $k$, $1 \le k \le n-1$, our streaming $O(n/k)$-pass algorithm computes {\em exact} single-source shortest paths from a designated root $r$ to all other vertices, using $O(n k)$ memory.
\end{theorem}

Consider now the scenario that we want to compute $s$-SSP, i.e., shortest paths from $s$ sources, for some $1 \le s \le n$.
We denote the sources $r_1,\ldots,r_s$. 

For a parameter $k$, we compute the hopset $G^{(k)}$ as described above, in $O(n/k)$ passes, using $O(n k)$ memory.
We then conduct Bellman-Ford from designated sources in $G \cup G^{(k)}$,  for $O(n/k)$ iterations, i.e., using $O(n/k)$ passes over the stream. (After each pass over the stream, we also scan again the hopset, stored in the local memory.) These passes use $O(n \cdot s)$ memory, i.e., $O(s)$ memory for every vertex, which it uses to store its $s$ distance estimates. This is in addition to the memory used to store the hopset itself. 

As a result, we obtain exact $S \times V$ distances in $G$, using $O(n (s+ k))$ memory, in $O(n/k)$ passes.
To obtain actual distances, we consider the $s$ SPTs $\tau_1,\ldots,\tau_s$ in $G \cup G^{(k)}$, rooted at the $s$ designated sources $r_1,\ldots,r_s$, respectively, which our algorithm has computed.
We process these trees (offline) one after another. We start by replacing every hopset edge in $\tau_1$ by an actual path (containing at most $k$ edges) in $G$. As a result we obtain a subgraph $T'_1$ with $O(n k)$ edges, such that for every vertex $v \in V$, we have 
$d_{T'_1}(r_1,v) = d_G(r_1,v)$. We then compute an SPT $T_1$ of $T'_1$ with respect to $r_1$, and erase $T'_1$ from the local memory.
Then we do the same with $\tau_2$, then with $\tau_3$, etc.

We summarize this argument below.

\begin{theorem}
\label{thm:stream_multi_src}
For any $n$-vertex weighted undirected graph $G = (V,E)$, and any parameter $k$, $1 \le k \le n-1$, our deterministic streaming $O(n/k)$-pass algorithm computes exact $s$-source shortest paths, using $O(n (k + s))$ memory.
\end{theorem}

A particularly useful setting of parameters here is $k=s$. Then we get $O(n/s)$ passes, $O(n \cdot s)$ memory, for exact $s$-SSSP.
Generally, for a fixed $s$, it makes sense only to use this theorem with $k \ge s$.

\subsection{Directed Graphs}
\label{sec:dir_stream}

In this section we show that the results of Theorems \ref{thm:stream_spt} and \ref{thm:stream_multi_src}
can be extended to directed graphs with negative edge weights, at the expense of losing a logarithmic factor in the number of passes, and by using randomization. 

We start with assuming that there are no cycles with negative weight in the graph. We will later show how to get rid of this assumption.

For a parameter $k$, $1 \le k \le n$, we sample every vertex $v \in V$ into the set $V'$ of virtual vertices u.a.r. with probability $k/n$.
We add the designated root vertex $r$ into $V'$. (We consider the single-source case first.)

We conduct B-F explorations from all the vertices of $V'$ in parallel, to depth $c \cdot {n \over k} \cdot \ln n$, for a sufficiently large constant $c$. Denote $\Gamma = c \cdot {n \over k} \cdot \ln n$. This involves $\Gamma$ passes over the stream of edges, and for every vertex $v \in V$, on each iteration $i \le \Gamma$, we store up to $|V'|$ current distance estimates $\{d_G^{(i)}(v',v) \mid v' \in V'\}$. 
Hence, we use expected $O(n \cdot k)$ memory for this step.

We obtain a virtual digraph $G' = (V',E')$,  defined by
$$E' = \{\langle u',v' \rangle \mid  v' \mbox{~~is~reachable~from~} u' \mbox{~~via~a~} \Gamma-\mbox{limited~path~in~} G\}~,$$
and with weight function $\omega'(\langle u',v'\rangle) = d_G^{(\Gamma)}(u',v')$.

\begin{lemma}
\label{lm:dir_dist}
Whp, for every $u',v' \in V'$, we have $d_{G'}(u',v') = d_G(u',v')$.
\end{lemma}
\proof
Since any $u'-v'$ path in $G'$ can be implemented by $u'-v'$ in $G$ of the same length, it follows that $d_{G'}(u',v') \ge d_G(u',v')$. 
In the opposite direction, let $P = P(u',v')$ be a shortest $u'-v'$ path in $G$ from the collection $\cP$ of shortest paths (see Section \ref{sec:hopset}), and denote $h = |P|$.
If $h \le \Gamma$, then there is an arc $\langle u',v' \rangle \in E'$ of weight $\omega'(\langle u',v' \rangle) = \omega(P(u',v')) = d_G(u',v')$, and so $d_{G'}(u',v') \le d_G(u',v')$. 

Otherwise, with high probability,  there are virtual vertices $u' = u'_0,u'_1,\ldots,u'_{q-1},u'_q = v'$ on the path $P$, such that the hop-distance in $P$ (and so in $G$ too) between every pair of these consecutive virtual vertices is at most $\Gamma$.
Hence all the arcs $\langle u'_0,u'_1 \rangle, \langle u'_1,u'_2\rangle, \ldots,\langle u'_{q-1},u'_q\rangle \in E'$, and for each index $i \in [0,q-1]$, we have $\omega'(\langle u'_i,u'_{i+1} \rangle) = d_G(u'_i,u'_{i+1})$.
Hence the path $\langle u'= u'_0,u'_1,\ldots,u'_q = v'\rangle$ is contained in $G'$, and its length is $\omega(P) = d_G(u',v')$. Hence $d_{G'}(u',v') \le d_G(u',v')$, whp.
\QED

Observe also that for every vertex $v \in V \setmns V'$, we store all the distances $d_G(v',v)$, such that $v$ is reachable from $v'$ via a $\Gamma$-limited path.

For a virtual vertex $u' \in V'$, and a vertex $v \in V \setmns V'$, which is not reachable from $u'$ via a $\Gamma$-limited path, note that, whp, the shortest $u'-v'$ path $P$ in $G$  contains a virtual vertex $v'$ within $\Gamma$ hops from $v$, and moreover, $d_{G'}(u',v) = d_G(u',v') + d_G(v',v)$.

To complete the algorithm, we compute offline all distances $r-v'$ in $G'$, for all $v' \in V'$. 
To store the virtual graph itself, we need only $O(k^2) = O(n k)$ memory.  Now, for every vertex $v$, we select the vertex $v' \in V'$ such that $v$ is reachable from $v'$ via a $\Gamma$-limited path, that minimizes $d_G(r,v') + d_G(v',v)$. By the above argument, this is  equal to $d_G(r,v)$.

If we are interested in distances from $s$ designated sources $r_1,\ldots,r_s$, then we use $k \ge s$, and include all these sources in $V'$. 
In the last step of the algorithm, instead of computing the distance from $r$ to $v$ (for every $v \in V$), we do it for every $r_i$, $i \in [s]$.
The number of passes and the memory requirement of the algorithm remain unchanged.

To retrieve the actual paths, we store for every vertex $v$ not just the distance estimate from each virtual vertex $v'$ whose B-F exploration reached $v$, but also the arc $\langle u,v \rangle \in E$ connecting $v$ to the parent $u$ through which the exploration of $v'$ reached $v$.
(This still requires an expected $O(n \cdot k)$ memory.)

Also, when computing an SPT $T'$ in $G'$ rooted at $r$ (or SPTs $T'_i$ rooted at $r_i$, for every $i \in [s]$, in the case of multiple sources), we store for every virtual vertex $v'$, its parent $u'$ in $T'$. Also, for every incoming arc $\langle u',v'\rangle \in E'$, the virtual vertex $v'$ stores its incoming $G$-parent $u \in V$. This enables us to retrieve the actual shortest path trees in $G$, while maintaining the same guarantees on space and the number of phases.

Consider now the case that negative-weight cycles may appear in the graph $G$. Consider the negative weight cycle $C = \langle u_0,u_1,\ldots,u_{q-1},u_q = u_0 \rangle$ in $G$ with minimum number of hops, reachable from the root $r$. (In the case of multiple sources,  the cycle needs to be reachable from one of the sources.)

If $q < \Gamma$, then let $v'$ denote the closest selected (aka virtual) vertex to a vertex of $C$. Suppose wlog that $d_G(v',C) = d_G(v',u_0)$. Then the distance estimate of $u_0$ will keep decreasing after $\Gamma$ iterations of the B-F from the selected vertices. Thus, to detect such cycles $C$, we modify the algorithm so that it will conduct $2 \Gamma$ iterations of  B-F, instead of $\Gamma$ ones.
However, we will only keep $\Gamma$-limited distance estimates. On the other hand, in the additional $\Gamma$ iterations we will check if an estimate of some vertex $v \in V$ decreases below $d^{(\Gamma)}(r,v)$. If this happens, the algorithm reports that the graph contains a negative-weight  cycle.

 Indeed, in the cycle $C$ as above, the distance estimate of $u_0$ will necessarily decrease in one of these additional $\Gamma$ iterations of B-F.  On the other hand, if the graph contains no negative-weight cycle, then as we have seen, whp, for every vertex $v \in V$, its distance estimate after $\Gamma$ iterations is already equal to $d_G(r,v)$, and thus it will not decrease any more. Hence, whp, no negative-weight cycle will be reported, if there is no such a cycle in $G$.

If $q \ge \Gamma$, then we replace the path $\langle u_0,\ldots,u_{q-1}\rangle$ in the cycle $C$ by the shortest $q$-limited path $P$ from $u_0$ to $u_{q-1}$, such that $P \in \cP$. (Note that this quantity is well-defined, even when negative-weight cycles are present.)
We obtain a possibly different negative-weight cycle $\tC = \langle u_0,u'_1,u'_2,\ldots,u'_{q-2},u_{q-1},u_q = u_0 \rangle$, with the same number of hops. (Recall that $C$ is the negative-weight cycle with minimum number of hops.) 

Moreover, whp, every $\Gamma$ hops in the cycle $\tC$ contain a selected vertex $v' \in V'$. There are two possibilities.
Either the entire cycle has length $q \le 2\Gamma$, and then it may contain just one selected vertex.  If this is the case, the estimate of $v'$ will decrease when we will conduct the additional $\Gamma$ iterations of B-F, and the algorithm will report that there is a negative-weight cycle.
On the other hand, if $q > 2\Gamma$, then, whp, the cycle contains at least two selected vertices $v'_1,v'_2 \in V'$, and moreover, the selected vertices $v'_1,v'_2,\ldots,v'_p$, for some $p \ge 2$, in $\tC$ form a cycle $C'$ in $G'$. (Because, whp, all hop-distances between consecutive selected vertices on the cycle $\tC$ are at most $\Gamma$.) It follows that there is a negative-weight cycle $C' = \langle v'_1,v'_2,\ldots,v'_p,v'_1\rangle$ in $G'$. This cycle will be detected offline by a B-F exploration in $G'$.

Hence, whp, if the graph $G$ contains a negative-weight cycle, then it will be detected by the algorithm, and on the other hand, when there is no negative-weight cycle, no distance estimate will decrease within the additional $\Gamma$ iterations of B-F, and thus the algorithm will not report that there is a negative-weight cycle.

We summarize this analysis below.

\begin{theorem}
\label{thm:dir_stream} 
For any $n$-vertex directed weighted graph $G = (V,E)$, with possibly negative edge weights, and any $s$ designated sources $r_1,\ldots,r_s$, and any parameter $k$, $s \le k \le n-1$, whp, our randomized algorithm computes exact $s$-sources shortest paths using $O({n \over k} \log n)$ passes and expected $O(n k)$ space. If there is a negative-weight cycle reachable from one of the sources,  the algorithm (whp) reports one, and, whp, when there is no negative-weight cycle, the algorithm reports that there is no such a cycle, and computes shortest paths.
\end{theorem}

\section*{Acknowledgements}

The author wishes to thank anonymous referees of STOC'17 conference for helpful remarks, and Keren Censor-Hillel, Sebastian Krinninger and Ofer Neiman for helpful discussions.

\def\APPlarge_band
{
\section{Large Bandwidth}
\label{sec:large_band}

In this section we analyze our algorithm in the $\CONGEST(b \log n)$ model, where $1 \le b \le n$ is the bandwidth parameter, i.e., when every edge can relay up to $b$ messages of size $O(\log n)$ each round. We start with the single-source case (Section \ref{sec:single_src_large_band}), and then analyze the case of multiple sources (Section \ref{sec:many_srcs_large_band}).

We remark that the benchmark exact SPT algorithm in the $\CONGEST(b \log n)$ model is still the Bellman-Ford, which requires $O(n)$ time for single source, and $O(n \lceil {s \over b} \rceil)$ for $s$ sources. Note that if really huge messages of $b = O(|E|)$ size are allowed, one can solve the problem (as well as any other problem) in $O(D)$ time by collecting the entire topology in a single vertex, solving the problem locally, and disseminating the solution. For more general bandwidth parameter $b$, the same solution works in $O(D + |E|/b)$ time, for the single-source case, and in $O(D + (|E| + ns)/b)$ time, for the case of $s$ sources. This solution is, however, generally very problematic, as it requires  heavy local computation.

\subsection{Single Source}
\label{sec:single_src_large_band}

Consider the construction of the $k$-shortcut hopset $\Gtagk$ via a B-F in $G$ to depth $O\left({{\log n} \over q} \cdot k\right)$, where every vertex $v$ forwards $k$ smallest estimates of its distances from vertices of $V'$ that it knows in every super-round. When the bandwidth is $b$ (words, i.e., $O(b \log n)$ bits), each super-round can be implemented in $\lceil k/b\rceil$ time, i.e., this step requires $O\left({{\log n} \over q} \cdot k\lceil {k \over b} \rceil\right)$ time.   

The upcast and pipelined broadcast over the BFS tree $\tau$ of $G$ to disseminate the hopset $\Gtagk$ can be now implemented in $O(D + {{nq k} \over b})$ time. (We disseminate $m = nqk$ messages over a tree of depth $D$, with edges of bandwidth $b$. See Lemma \ref{lm:upcast} in Appendix \ref{sec:upcast}.) 

The B-F in $G' \cup \Gtagk$ is conducted for $h' = O\left({{nq } \over k}\right)$ iterations. The first part of every iteration involves dissemination of $O(n \cdot q)$ estimates of distances of virtual vertices from the designated root vertex $r$ over the BFS tree $\tau$. This step requires $O(D + \lceil {{nq} \over b} \rceil)$ time. The second part of every iteration is a B-F in $G$ to depth $O\left({{\log n} \over q}\right)$ from vertices of $V'$. Each vertex forwards just one single smallest estimate of distance from $r$ that it knows. Here the larger bandwidth does not help, and this step  still requires $O\left({{\log n} \over q}\right)$ time, exactly as in the unit-bandwidth case.

Hence the overall running time is 
\begin{equation}
\label{eq:T_ss}
 T ~=~ O\left(\left(D +\lceil {{nq} \over b} \rceil + {{\log n} \over q}\right) \cdot {{nq} \over k} + {{\log n} \over q} \cdot k \cdot \lceil {k \over b} \rceil + {{nqk} \over b}\right)~.
\end{equation}
In this expression we always require $k \le nq$. We will also have $b \le nq$, because the last term $nqk/b$ is typically dominated by other terms.

For the range of small $D$, i.e., $D = O\left(\sqrt{{n \log n} \over b}\right)$, we will use two settings of the parameters, depending on the value of $b$. In both settings we will set $q = \sqrt{{b \log n} \over n}$. When $b$ is relatively small, i.e., $b \le (n \log n)^{1/3}$, we set $k = (n \log n)^{1/6} \cdot b^{1/2}$. We have $b \le k = (n \log n)^{1/6} \cdot b^{1/2}$, and $k = (n \log n)^{1/6} \cdot b^{1/2} \le nq = \sqrt{n \log n \cdot b}$. The running time of the algorithm becomes 
\begin{equation}
\label{eq:time_smallb}
T = O\left(({{nq} \over b} + {{\log n} \over q}) \cdot {{nq} \over k} + {{\log n} \over q} \cdot {{k^2} \over b} + {{nqk} \over b}\right) ~=~
O\left({{(n \log n)^{5/6} } \over {b^{1/2}}}\right)~.
\end{equation}
(Note that ${{nq} \over b} \cdot k = \sqrt{{n \log n} \over b} \cdot (n \log n)^{1/6} \cdot b^{1/2} = (n \log n)^{2/3} \le {{(n \log n)^{5/6}} \over {b^{1/2}}}$, for $b \le (n \log n)^{1/3}$.)

When $b \ge (n \log n)^{1/3}$, we set $k = \sqrt{nq} = (n \log n \cdot b)^{1/4}$. 
Here $k \le b \le nq$. Indeed, $(n \log  n \cdot b)^{1/4} \le b \le \sqrt{n \log n \cdot b}$, for this range of $b$. (Recall that we assume that $b$ is at most linear in $n$.) Hence, by (\ref{eq:T_ss}),  the running time is (as $\lceil {k \over b} \rceil =1$)
$$
T ~=~ O\left(\left({{nq} \over b} + {{\log n} \over q}\right) \cdot {{nq} \over k} + {{\log n} \over q} \cdot k + {{nq k} \over b}\right)~.
$$
Then
\begin{equation}
\label{eq:time_largeb}
T ~=~O\left({{(n\log n)^{3/4}} \over {b^{1/4}}}\right)~.
\end{equation}
Observe that the two bounds (\ref{eq:time_smallb}) and (\ref{eq:time_largeb}) agree when $b = \Theta((n \log n)^{1/3})$, and are equal to $O((n \log n)^{2/3})$. Also, for $b \approx n$, the bound (\ref{eq:time_largeb}) gives time $\tO(\sqrt{n})$. 

For the case of large $D$, i.e., $D \ge \sqrt{{n \log n} \over b}$, we will assume $b \le k \le nq$. 
Assuming $D \ge \max\{ \lceil {{nq} \over b} \rceil, {{\log n} \over q} \}$, by (\ref{eq:T_ss}), 
we then have
$$T = O\left(D \cdot {{nq} \over k} + {{\log n} \over q} \cdot {{k^2} \over b} + {{nqk} \over b}\right)~.$$
    We set $q = {{\log n} \over D}$, $k = \left({{n \log n \cdot b} \over D}\right)^{1/3}$. 
Note that it indeed holds that $D \ge \max\{ \lceil {{nq} \over b} \rceil, {{\log n} \over q} \}$.
The condition $k \le nq$ implies $D \le {{n \log n} \over {b^{1/2}}}$, i.e., $b \le \left({{n \log n} \over D}\right)^2$. 
The condition $b \le k$ implies $b \le \left({{n \log n} \over D}\right)^{1/2}$, i.e.,  $D \le {{n \log n} \over {b^2}}$.
The time becomes 
\begin{equation}
\label{eq:time_largeD}
T ~=~ O\left(\left({D \over b}\right)^{1/3} (n \log n)^{2/3}\right)~,
\end{equation}
assuming $b \le (n \log n)^{1/3}$.
(Observe that $nqk/b = ((n \log n)/D)^{4/3}/b^{2/3} \le (D/b)^{1/3} (n \log n)^{2/3}$ iff $D \ge (n \log n)^{2/5}/b^{1/5}$. However, 
$D \ge \sqrt{{n \log n} \over b}$, and $\sqrt{{n \log n} \over b} \ge {{(n \log n)^{2/5}} \over {b^{1/5}}}$, for $b \le (n \log n)^{1/3}$.)
Also, the conditions $D \ge \sqrt{{n \log n} \over b}$ and $b \le \left({{n\log n} \over D}\right)^{1/2}$ imply 
$D \ge (n \log n \cdot D)^{1/4}$, i.e., $ D \ge (n \log n)^{1/3}$. 

We summarize this analysis in the next theorem.

\begin{theorem}
\label{thm:single_src_large_band}
Our algorithm computes single-source exact SPT in $\CONGEST(b \log n)$ model in time:
\begin{enumerate}
\item
\label{item:smallb_smallD}
If $b \le (n \log n)^{1/3}$ and $D = O(\sqrt{{n \log n} \over b})$, then in $O\left({{(n \log n)^{5/6}} \over {b^{1/2}}}\right)$ time.
\item
\label{item:largeb_smallD}
If $b \ge (n \log n)^{1/3}$ and $D =  O(\sqrt{{n \log n} \over b})$, then in $O(\left({{(n \log n)^{3/4}} \over {b^{1/4}}}\right)$ time.
In particular, for $ b = \Theta(n)$, $D =  O(\sqrt{\log n})$, the running time is $\tO(\sqrt{n})$. 
\item
\label{item:smallband_largeD}
If  $b \le (n \log n)^{1/3}$ and $D \ge (n \log n)^{1/3}$, and also $\sqrt{{n \log n} \over b} \le D \le {{n \log n} \over {b^2}}$, then in 
$O((D/b)^{1/3} (n \log n)^{2/3})$ time. In particular, for $b \approx D \approx (n \log n)^{1/3}$, the time is $O((n \log n)^{2/3})$.
\end{enumerate}
\end{theorem}

So for $D = n/\alpha$, for a parameter $\alpha$, we use $b = (\alpha \log n)^{1/2}$, and get running time $O\left({{n \cdot \log^{1/2} n} \over {\alpha^{1/2}}}\right)$, i.e., it is sublinear in $n$ for $\alpha = \omega(\log n)$ (that is, for almost entire range of $D$, specifically, 
 $D = o(n/\log n)$). 


The bound \ref{item:smallb_smallD} of Theorem \ref{thm:single_src_large_band} is always better than the (linear) bound of the Bellman-Ford algorithm. Comparing it will the topology-collecting algorithm (that requires $O(D + |E|/b)$) time, our algorithm has smaller running time whenever $|E| = \omega(n \log n)$. When $|E| = O(n \log n)$, our algorithm has still smaller running time than the topology-collecting  algorithm if $b = o(n^{1/3}/\log^{5/3} n)$. Only if the graph is very sparse ($|E| = o(n \log n)$) and the bandwidth $b$ satisfies $\Omega(n^{1/3}/\log^{5/3} n)
 = b = O((n \log n)^{1/3})$, the topology-collecting algorithm has smaller running time than our algorithm, by a polylogarithmic in $n$ factor.
(But, of course, our algorithm avoids heavy local computations, while the topology-collecting algorithm heavily relies on them.)

The bound \ref{item:largeb_smallD} of Theorem \ref{thm:single_src_large_band} is also always better than the linear bound of Bellman-Ford.
 Here our running time loses to the topology-collecting algorithm only when $b = \omega\left({{|E|^{4/3}} \over {n \log n}}\right)$ and 
$D \le \sqrt{{n \log n} \over b} = {{n \log n} \over {|E|^{2/3}}}$.  In particular, if $|E| = \omega((n \log n)^{3/2})$, then our bound is always better than that of the topology-collecting algorithm. 

The bound \ref{item:smallband_largeD} of Theorem \ref{thm:single_src_large_band} is sublinear in $n$ as long as $D \le {n \over {\log^2 n}} \cdot b$, i.e., also almost in the entire range of parameters. This bound is also better than that of the topology-collecting algorithm whenever $|E| = \omega(n \log n)$. Even when the graph is very sparse (i.e., $|E| = O(n \log n)$), our bound loses (by at most a polylogarithmic in $n$ factor)  to that of the topology-collecting algorithm only when the diameter $D$ gets close (within a  polylogarithmic in $n$  factor) to its upper bound ${{n \log n} \over {b^2}}$.

\subsection{Multiple Sources}
\label{sec:many_srcs_large_band}

As we have seen, the construction of the $k$-shortcut hopset $\Gtagk$ requires $O\left({{\log n} \over q} \cdot k \cdot \lceil {k \over b} \rceil\right)$ time in the $\CONGEST(b \log n)$ model. The upcast and pipelined broadcast of its edges over the BFS tree $\tau$ of $G$ are performed within an additional $O(D + {{n q k} \over b})$ time. 

Like in Section \ref{sec:mult_src}, we assume that $s \le nq$. 

After the hopset $\Gtagk$ was constructed, the algorithm conducts a B-F in $G' \cup \Gtagk$ for $h' = O(nq/k)$ iterations. (So $k \le nq$ is another constraint.)
 The first part of each iteration of the B-F involves collecting and disseminating $O(nq s)$ estimates over $\tau$, i.e., it requires $O(D + {{nqs} \over b})$ time.

In the second part of each iteration, a B-F in $G$ to depth $O((\log n)/q)$ is conducted. Here in each step the algorithm needs to relay up to $s$ estimates. When the bandwidth is $b$, each such a step requires $O(\lceil {s \over b} \rceil)$ rounds. Hence, overall, the second part of each iteration requires $O\left({{\log n} \over q} \cdot \lceil{s \over b} \rceil\right)$ time.

Thus, each iteration of the B-F in $G' \cup \Gtagk$ is executed within $O(D + {{nqs} \over b} + {{\log n} \over q} \cdot \lceil{s \over b} \rceil)$ time.  Hence the entire B-F in $G' \cup \Gtagk$ requires 
$O((D + {{nqs } \over b} + {{\log n} \over q} \cdot \lceil {s \over b} \rceil) \cdot {{nq} \over k})$ time. 

The final step of the algorithm, where vertices of $V \setmns V'$ learn their distances, requires $O\left({{\log n} \over q} \cdot \lceil {s \over b} \rceil\right)$ time. This term is dominated by the running times of other steps of the algorithm.

Hence the total running time of the algorithm is given by
\begin{equation}
\label{eq:time_many_srcs_large_band}
T~=~ O\left({{\log n} \over q} \cdot k \cdot \lceil {k \over b} \rceil + D + {{nqk} \over b} + (D + {{nqs} \over b} + {{\log n } \over q} \cdot \lceil {s \over b} \rceil)\cdot {{nq } \over k}\right)~.
\end{equation}
To analyze this expression, we consider a number of regimes of parameters.
\inline Case 1: Here we restrict the parameters to satisfy $s \le b \le k (\le nq)$. (The parenthesized inequality always holds.)
Then
\begin{equation}
\label{eq:time_fst_case}
T ~=~ O\left((D + {{nqs} \over b} + {{\log n} \over q}) \cdot {{nq} \over k} + {{\log n} \over q} \cdot {{k^2} \over b} + {{nqk} \over b}\right)~.
\end{equation}
Consider first the subcase of small diameter, i.e., $D \le \sqrt{{s \over b} n \log n}$.  Set $q = \sqrt{{b \log n} \over {ns}}$, $k = {{(n \log n)^{1/6} \cdot b^{1/2}} \over {s^{1/6}}}$. It follows that 
\begin{equation}
\label{eq:fst_case_smallD}
T ~=~ O\left({{(n\log n)^{5/6}} \over {b^{1/2}}} \cdot s^{1/6}\right)~.
\end{equation}
This bound applies when $s \le (n \log n)^{1/4}$, $s \le b \le \left({{n \log n} \over s}\right)^{1/3}$. (The right-hand inequality follows from $b \le k$.)
Note that for $s = b \le (n \log n)^{1/4}$, $D \le \sqrt{n \log n}$, we have here $T = O\left({{(n \log n)^{5/6}} \over {b^{1/3}}}\right)$. 

For large diameter, i.e., $D \ge \max\{\sqrt{{n s \log n} \over b}, (n \log n)^{1/3} \cdot s^{2/3}\}$, we set $q = {{\log n} \over D}$, $k = \left({{n \log n \cdot b} \over D}\right)^{1/3}$. 
Note that 
$${{nq k} \over b} ~=~ {n \over b} {{\log n} \over D} \cdot \left({{n \log n \cdot b} \over D}\right)^{1/3} ~=~ \left({{n \log n} \over D}\right)^{4/3}/b^{2/3}~.
$$
This expression is dominated by $(n \log n)^{2/3} (D/b)^{1/3}$, as long as $D \ge {{(n \log n)^{2/5} } \over {b^{1/5}}}$. Since 
$D \ge \sqrt{{n \log n \cdot s} \over b}$, it is sufficient to show that $s^{5/6} (n \log n)^{1/3} \ge b$. But we also have $b \le \left({{n \log n} \over D}\right)^{1/2}$. (This follows from $b \le k = \left({{n \log n \cdot b} \over D}\right)^{1/3}$.)
Hence 
$$b ~\le ~
\left({{n \log n \cdot \sqrt{b}} \over {\sqrt{n \log n \cdot s}}}\right)  ~=~ 
\left({{n \log n \cdot b} \over s}\right)^{1/4}~,$$ and so $b \le \left({{n \log n} \over s}\right)^{1/3}$.
Thus, $s^{5/6} (n \log n)^{1/3} \ge ((n\log n)/s)^{1/3} \ge b$ follows.

We conclude that 
\begin{equation}
\label{eq:fst_case_largeD}
T ~=~ O((n \log n)^{2/3} (D/b)^{1/3})~,
\end{equation}
subject to $D \ge \max\{\sqrt{{n \log n \cdot s} \over b},(n \log n)^{1/3} s^{2/3}\}$. 
(The inequality $D \ge (n \log n)^{1/3} \cdot s^{2/3}$ is because $D \ge \sqrt{{n \log n \cdot s} \over b}$ and $b \le k = \left({{n b \log n} \over D}\right)^{1/3}$ imply $b \le \left({{n \log n} \over D}\right)^{1/2}$.)
We also need $s \le b \le \left({{n \log n} \over D}\right)^{1/2}$ to hold, for (\ref{eq:fst_case_largeD}) to be valid.

So given $b$ and $D$, the bound (\ref{eq:fst_case_largeD}) holds for $s$ sources, as long as $s \le \min \{b, {{D^2 b} \over {n \log n}}\}$. This bound generalizes Theorem \ref{thm:single_src_large_band}, bound \ref{item:smallband_largeD}.

Another regime of parameters which gives rise to meaningful bounds is the following one.

\inline Case 2: $b \le s$, $b \le k (\le nq)$.

Here the running time (see (\ref{eq:time_many_srcs_large_band}) becomes
\begin{equation}
T ~=~ O\left(\left(D + {{nq s} \over b} + {{\log n} \over q} \cdot {s \over b}\right) \cdot {{nq} \over k} + {{\log n} \over q} \cdot {{k^2}  \over b} + {{nq k} \over b}\right)~.
\end{equation}
First, we consider the subcase of small diameter, i.e., $D \le {s \over b} \sqrt{n \log n}$. We set $q = \sqrt{{\log n} \over n}$, $k = (n \log n)^{1/6} \cdot s^{1/3}$, and obtain 
\begin{eqnarray}
\nonumber
T & = & O(\sqrt{n \log n} \cdot {s \over b} {{\sqrt{n \log n} } \over {(n \log n)^{1/6} \cdot s^{1/3}}} +
\sqrt{n \log n} \cdot {1 \over b} \cdot (n \log n)^{1/3} \cdot s^{2/3} + {{\sqrt{n \log n}} \over b} (n \log n)^{1/6} \cdot s^{1/3}) \\
\nonumber
& = & O\left((n\log n)^{5/6} \cdot {{s^{2/3}} \over b} + (n \log n)^{2/3} \cdot {{s^{1/3}} \over b}\right) \\
\label{eq:scnd_case_smallD}
& = & O\left((n\log n)^{5/6} \cdot {{s^{2/3}} \over b}\right)~. 
\end{eqnarray}
This bound applies for $b \le \min\{s, (k = ) (n \log n)^{1/6} \cdot s^{1/3}\}$, and $D \le {s \over b} \sqrt{n \log n}$. 

For $s = b$, this bound agrees with the bound (\ref{eq:fst_case_smallD}), i.e., both bounds give $T = O((n \log n)^{5/6} /b^{1/3})$.
Both these bounds (\ref{eq:fst_case_smallD}) and (\ref{eq:scnd_case_smallD}) are non-trivial, i.e., they are smaller than $O(\min \{n \lceil {s \over b}\rceil,D + {{|E|+ ns} \over b}\}$ almost in the entire range of parameters.
See the discussion following Theorem \ref{thm:large_band_smallD}. 


Now we consider the case of large diameter, i.e., $D \ge {s \over b} \sqrt{n \log n}$. (We also have $b \le s$, $b \le k (\le nq)$.) 
Here we have 
$$T ~=~ O(D \cdot {{nq} \over k} + {{\log n} \over q} \cdot {{k^2} \over b} + {{nq k} \over b})~.$$
We set $q = \sqrt{{\log n} \over n}$, $k = (Db)^{1/3}$, and get 
\begin{equation}
\label{eq:T_one_bound}
T ~=~ T_1 ~=~ O\left({{D^{2/3} \sqrt{n \log n} } \over {b^{1/3}}}\right)~,
\end{equation}
 subject to $b \le s \le b \cdot {D \over {\sqrt{n \log n}}}$, and 
$\max\{ {s \over b} \sqrt{n \log n}, b^2\} \le D \le {{(n \log n)^{3/2}} \over b}$. 

Another option (still for $b \le s$, $b \le k \le nq$, $D \ge {s \over b} \sqrt{n \log n}$) is to set $q = {{s \log n} \over {D \cdot b}}$.
This guarantees ${{\log n} \over q} \ge n q$, i.e., $T = O\left(D \cdot {{nq} \over k} + {{\log n} \over q} \cdot {{k^2} \over b}\right)$.
(The third term ${{nqk} \over b}$ is dominated by ${{\log n} \over q} \cdot {{k^2} \over b}$.) Set $k = \left({{n \log n} \over {D b}}\right)^{1/3} \cdot s^{2/3}$.
We get
\begin{equation}
\label{eq:T_two_bound}
T ~=~ T_2 ~=~ O\left((D s)^{1/3} \left({{n \log n} \over b}\right)^{2/3}\right)~.
\end{equation}
The condition $b \le k$ gives rise to $b \le \min\{s, s^{1/2} \left({{n \log n} \over D}\right)^{1/4}\}$. 
The condition $k \le nq$ (together with the above) gives rise to ${s \over b} \sqrt{n \log n} \le D \le n \log n \cdot {{\sqrt{s}} \over b}$.

Note that $T_2 \le T_1$ whenever $D \ge {s \over b} \sqrt{n \log n}$, i.e., whenever both bounds apply, $T_2$ is better than $T_1$. Hence $T_1$ is meaningful only for $D > n \log n {{\sqrt{s}} \over b}$, or for $s \ge b \ge s^{1/2}\left({{n \log n} \over D}\right)^{1/4}$. 
Moreover, in fact, if $b \le  s^{1/2}\left({{n \log n} \over D}\right)^{1/4}$, 
then $D \ge n \log n {{\sqrt{s}} \over b}$ (or else $T_2$ is applicable).
 But this implies $b \le b^{1/4} \cdot s^{3/8}$, i.e., $b \le \sqrt{s}$. Then, however, the condition $D \ge {{\sqrt{s}} \over b} n \log n$ cannot hold.  So for $T_1$ to be meaningful, it must hold that $b \ge s^{1/2}\left({{n \log n} \over D}\right)^{1/4}$.
However, using $T_2$ with $b = s^{1/2} \left({{n \log n} \over D}\right)^{1/4}$ (even when $b$ is, in fact, larger than that value) gives rise to
$$T_2 ~=~ O\left((Ds)^{1/3} \left({{(n \log n)^{3/4} \cdot D^{1/4}} \over {s^{1/2}}} \right)^{2/3} \right) ~=~ O(\sqrt{D n \log n})~.$$
On the other hand, $T_1$ applies for $b^2 \le D$, and in this range $D^{2/3} \sqrt{n \log n}/b^{1/3} \ge \sqrt{D n \log n}$. Hence $T_2$ is never worth than $T_1$.

Yet another bound for the case $b \le s$ and large diameter $D$ arises when we consider the case $s \ge b \ge k$, $nq \ge k$.
Here, by (\ref{eq:time_many_srcs_large_band}),  the running time behaves as 
$$T ~=~ O\left(\left(D + {s \over b}(nq + {{\log n} \over q})\right){{nq} \over k} + {{\log n} \over q} \cdot k + {{nq k} \over b}\right)~.$$
When $D$ is large, i.e., $D \ge \max\{nq,{{\log n} \over q}\} \ge {s \over b} \sqrt{n \log n}$, the bound becomes 
\begin{equation}
\label{eq:T_largeD}
T ~=~ O\left(D \cdot {{nq} \over k} + {{\log n} \over q} \cdot k + {{nq k} \over b}\right)~.
\end{equation}
We set $q = \sqrt{{\log n} \over n}$, $k = \sqrt{D}$, and obtain $T = T_3 = O(\sqrt{n \log n \cdot D})$, subject to 
$s^2 \ge b^2 \ge D \ge {s \over b} \sqrt{n \log n}$. (This follows from $s \ge b \ge k = \sqrt{D}$.)
It also follows that $b \ge s^{1/3} (n \log n)^{1/6}$, and 
$s \ge \max\{b, {{\sqrt{n \log n} } \over b}\} \ge (n \log n)^{1/4}$. 

Recall that $T_1$ applies only when $D \ge b^2$, i.e., in the complimentary range. 
Indeed, for $D = b^2$, we have $T_1, T_3= O(\sqrt{D n \log n})$. 
\inline Case 3 (Larger bandwidth):
Here we consider the case $b \ge s$, $b \ge k$, $nq \ge k$.

First, we consider the subcase of small diameter, i.e., $D \le \sqrt{{s \over b} n \log n}$. We set $q = \sqrt{{b \log n} \over {ns}}$, 
$k = \left({{n \log n \cdot b} \over s}\right)^{1/4}$, and, by (\ref{eq:time_many_srcs_large_band}), obtain running time
\begin{eqnarray}
\label{eq:T_smallD}
T & =& O\left(\left(D + {{nqs} \over b} + {{\log n} \over q}\right) \cdot {{nq} \over k}  + {{\log n} \over q} \cdot k + {{nq k} \over b}\right) \\
\nonumber
& = & O\left(\left({{nqs} \over b} + {{\log n} \over q}\right) \cdot {{nq} \over k}  + {{\log n} \over q} \cdot k + {{nq k} \over b}\right) \\
\nonumber
& =&  O\left(\sqrt{{n s \log n} \over b} \cdot \sqrt{{n \log n \cdot b} \over s} \cdot \left({s \over {n \log n} \cdot b}\right)^{1/4}\right)
\\
\nonumber
&+& O\left(
\sqrt{{n s \log n }  \over b} \cdot \left({{n \cdot \log n \cdot b} \over s}\right)^{1/4} + \sqrt{{n \log n \cdot b} \over s} \cdot {1 \over b} \left({{n \log n \cdot b} \over s}\right)^{1/4}\right) \\
\label{eq:three_stars}
& = & O\left((n \log n)^{3/4} \cdot \left({s \over b}\right)^{1/4}\right)~.
\end{eqnarray}
The condition $b \ge k$ implies $b \ge \max\{ s, \left({{n \log n} \over s}\right)^{1/3}\} \ge (n \log n)^{1/4}$.  The condition $k \le nq$ always holds. We always have $T = O((n \log n)^{3/4})$ in this range.

For $s = O(1)$, and $b = (n \log n)^{1-\eps}$, for a small constant $\eps > 0$, we obtain $D \le \sqrt{{s \over b} \cdot n \log n} = (n \log n)^{\eps/2}$, and $b \ge \max\{1, (n \log n)^{1/3}\} = (n \log n)^{1/3}$.
Here $T = (n \log n)^{1/2 + \eps/4}$. With $b = n/\polylog(n)$, we can have $T =\tO(\sqrt{n})$, for $D \le \polylog(n)$, $s = O(1)$. 

For $s = b = (n \log n)^{1/4}$, we get here $T = (n \log n)^{3/4}$, for $D \le \sqrt{n \log n}$.

For $s = 1$, we obtain here $T = O\left({{(n \log n)^{3/4}} \over {b^{1/4}}}\right)$, for $b \ge (n \log n)^{1/3}$. (This is the bound
\ref{item:largeb_smallD} of Theorem \ref{thm:single_src_large_band}.)
Formerly (see equation (\ref{eq:time_smallb}), we had $T  = O\left({{(n \log n)^{5/6}} \over {b^{1/2}}}\right)$, for $b \le (n \log n)^{1/3}$. 
(Both bounds apply for $D \le \sqrt{(n \log n)/ b}$.)
These bounds agree at $b = (n \log n)^{1/3}$, and give $T = O((n \log n)^{2/3})$. 

Now, consider the case of large $D$, i.e., $D \ge \max\{{{nqs} \over b},{{\log n} \over q}\} \ge  \sqrt{{s \over b} \cdot n \log n}$. First, we set $q$ so that $D \ge {{nq s} \over b} \ge {{\log n} \over q}$. Specifically, $q = \sqrt{{b \log n} \over {n s}}$, $k = \sqrt{{Db} \over s}$. 
Then , by (\ref{eq:T_smallD}), 
\begin{equation}
T ~=~ O\left(D \cdot {{nq} \over k} + {{\log n} \over q} \cdot k + {{nq k} \over b}\right) 
~ =~  O(\sqrt{D n \log n})~.
\end{equation}
For this bound to hold, we need $bs \ge D \ge \sqrt{{s \over b} \cdot n \log n}$, $b \ge s$. 
(The inequality $bs \ge D$ follows from $b \ge k = \sqrt{{Db} \over s}$.) This also means that $b^2 \ge bs \ge D$, i.e., $b \ge \sqrt{D}$. 
(Recall that we also have the bound $T = T_3 = O(\sqrt{D n \log n})$, for $s^2 \ge b^2 \ge D \ge {s \over b} \sqrt{n \log n}$. See (\ref{eq:T_largeD}).)

Another option is to set $q = b \sqrt{{\log n} \over {D n}}$, $k = b$. (We are still in the regime $b \ge s$.) 
Then
\begin{eqnarray}
\label{eq:time_Dn}
T & = & O\left(D \cdot {{nq} \over k} + {{\log n} \over q} \cdot k + {{nq k} \over b}\right) \\
\nonumber
& = & O\left(\sqrt{Dn \log n} + \sqrt{{n \log n} \over D} \cdot b\right)~.
\end{eqnarray}
Assuming $D \ge b$, the latter expression is $O(\sqrt{D n \log n})$ (subject to $D \ge b \ge s$). We also require $q = \Omega\left({{\log n} \over n}\right)$, and so $b \ge \sqrt{{D \log n} \over n}$, i.e., ${{n b^2} \over {\log n}} \ge D \ge b \ge s$. 
In addition, the conditions $D \ge {{\log n} \over q}$, $D \ge {{nq s} \over b}$ (under which (\ref{eq:time_Dn}) applies) imply 
$D \ge \max\{{{n \log n} \over {b^2}}, s^{2/3} \cdot (n \log n)^{1/3}\}$. 

Hence the bound $T = O(\sqrt{D n \log n})$ holds in three ranges:
\begin{enumerate}
\item
For $s \ge b \ge \sqrt{D}$, $D \ge {s \over b} \sqrt{n \log n}$.
\item
For $b^2 \ge bs \ge D \ge \sqrt{{s \over b} \cdot n \log n}$, $b \ge s$.
\item
For ${{n b^2} \over {\log n}} \ge D \ge \max\{ {{n \log n} \over {b^2}}, s^{2/3} (n \log n)^{1/3}\}$, $b \ge s$. 
\end{enumerate}

The next theorem summarizes the bounds that we have for the case of relatively small $D$.
\begin{theorem}
\label{thm:large_band_smallD}
Our algorithm computes exact shortest paths for $S \times V$, $|S| = s$, $\CONGEST(b \log n)$ model for the case of relatively small diameter $D$ in running time $T$ given by: \\
When $D \le \sqrt{{s \over b} n \log n}$, then
\begin{enumerate}
\item
\label{item:T_one}
$T ~=~ O\left({{(n \log n)^{5/6}} \over {b^{1/2}}} \cdot s^{1/6}\right)$, for $s \le O((n \log n)^{1/4})$, $s \le b \le \left({{n \log n} \over s}\right)^{1/3}$ (see (\ref{eq:fst_case_smallD})); and
\item
\label{item:T_three}
$T ~= ~ O((n \log n)^{3/4} \cdot (s/b)^{1/4})$,  for $b \ge \max\{s, ((n  \log n)/s)^{1/3}\} \ge (n \log n)^{1/4}$ (see (\ref{eq:three_stars})).
\end{enumerate}
When $D \le {s \over b} \cdot \sqrt{n \log n}$, and $b \le \min\{s,s^{1/3} (n \log n)^{1/6}\}$, then (see (\ref{eq:scnd_case_smallD}))
\begin{equation}
\label{eq:T_two}
T ~=~ O\left((n \log n)^{5/6} \cdot {{s^{2/3}} \over b}\right)~.
\end{equation}
\end{theorem}

Observe that for $b = s$, the bounds \ref{item:T_one} (of Theorem \ref{thm:large_band_smallD}) and (\ref{eq:T_two}) agree, and give 
$$T ~=~ O\left((n \log n)^{5/6} \cdot {1 \over {s^{1/3}}}\right) ~=~ O\left((n \log n)^{5/6} \cdot {1 \over {b^{1/3}}}\right)~.$$

The bound (\ref{eq:T_two}) is non-trivial (i.e., smaller than $O(\min\{n \lceil {s \over b} \rceil, D + {{ns+ |E|} \over b}\})$ in the entire range 
of parameters. 

Even if the graph is relatively sparse, i.e., 
$|E| = O(ns)$, still
the bound  \ref{item:T_one} of Theorem \ref{thm:large_band_smallD}   is non-trivial (i.e., it is $o(ns/b)$)  if $s = \omega(\log n)$, or (more generally) if $b = o\left({{n^{1/3} \cdot s^{5/3}} \over {\log^{5/3} n}}\right)$.
In other words, it loses to the topology-collecting algorithm only if $\omega\left({{n^{1/3} \cdot s^{5/3}} \over {\log^{5/3} n}}\right) = b = O\left( \left({{n \log n} \over s}\right)^{1/3}\right)$.
 For $\omega(ns) = |E| = o(n(s+b))$, the range in which this bound is non-trivial is even larger than that.
If $|E| = \Omega(n(s+ b))$, then the trivial algorithm requires at least $\Omega(n \lceil s/b \rceil) = \Omega(n)$ time, while the bound
  \ref{item:T_one}  is always sublinear. (Recall that $s \le (n \log n)^{1/4}$.) Hence for relatively dense graphs, this bound
  is always non-trivial.

The situation is similar with respect to bound \ref{item:T_three}. 
Even if the graph is relatively sparse, i.e., $|E| = O(ns)$, still the bound \ref{item:T_three} of Theorem \ref{thm:large_band_smallD} is non-trivial for $s \cdot {{n^{1/3}} \over {\log n} } \ge b \ge \max\{s, \left({{n \log n} \over s}\right)^{1/3}\}$. 
For $\omega(ns) = |E| = o(n (s+b))$, the range in which this bound is non-trivial is even larger than that. 
If $|E| = \Omega(n(s+ b))$, then the trivial algorithm requires at least $\Omega(n \lceil s/b \rceil) = \Omega(n)$ time, while the bound
  \ref{item:T_three}  is always sublinear. (Recall that $b \ge s$.) Hence for relatively dense graphs, this bound
  is always non-trivial.

The bound \ref{item:T_one} of Theorem \ref{thm:large_band_smallD}  generalizes the single-source bound  
\ref{item:smallband_largeD} 
of Theorem \ref{thm:single_src_large_band}. The bound \ref{item:T_three} of Theorem  \ref{thm:large_band_smallD} generalizes the single-source bound \ref{item:largeb_smallD} of Theorem \ref{thm:single_src_large_band}.
The bound (\ref{eq:T_two}) generalizes the multi-source unit-bandwidth bound (for small diameter) of Theorem \ref{thm:few_srcs}. 

In the next theorem we summarize the bounds that we have for the case of relatively large $D$.

\begin{theorem}
\label{thm:large_band_largeD}
Our algorithm computes exact shortest paths for pairs $S \times V$, $|S| = s$,  in $\CONGEST(b \log n)$ model for the case of relatively large diameter $D$ in running time $T$ given by: 
\begin{enumerate}
\item
\label{item:LargeD_one}
$T_0 = T = O((n \log n)^{2/3} (D/b)^{1/3})$, subject to 
$D \ge max\{ \sqrt{{s \over b} n \log n}, (n \log n)^{1/3} \cdot s^{2/3}\}$, $s \le b \le \left({{n \log n} \over D}\right)^{1/2}$.
(Equivalently, given $D,b$, the bound holds for $s \le \min\{ b, {{D^2 b} \over {n \log n}}\}$. See inequality (\ref{eq:fst_case_largeD}).)
\item
\label{item:LargeD_three}
$T = T_2 = O\left((Ds)^{1/3} \cdot \left({{n \log n} \over b}\right)^{2/3}\right)$, subject to 
$b \le \min\{ s, s^{1/2} \left({{n \log n} \over D}\right)^{1/4}\}$, 
${s \over b} \cdot \sqrt{n \log n} \le D \le n \log n \cdot {{\sqrt{s}} \over b}$.
Note that $T_2$ is better than $T_1$ whenever $T_2$ is applicable, and thus $T_1$ is meaningful only when $s \ge b \ge s^{1/2} ((n \log n)/D)^{1/4}$.  (See inequality (\ref{eq:T_two_bound}).) 
\item
\label{item:LargeDn}
The bound $T = T_3 = O(\sqrt{D n \log n})$, which applies in the three following domains:
\begin{enumerate}
\item
\label{subitem_one}
For $s \ge b \ge \sqrt{D}$, $D \ge {s \over b} \sqrt{n \log n}$.
\item
\label{subitem_two}
For $sb \ge D \ge \sqrt{{s \over b} \cdot n \log n}$, $b \ge s$.
\item
\label{subitem_three}
For $D \ge \max\{ {{n \log n} \over {b^2}}, s^{2/3} (n \log n)^{1/3}\}$, $b \ge s$. 
\end{enumerate}
\end{enumerate}
\end{theorem}
The bounds \ref{item:LargeD_one},   \ref{subitem_two}  and \ref{subitem_three} of Theorem \ref{thm:large_band_largeD} apply for $s \le b$.  Observe that in \ref{item:LargeD_one} and  \ref{subitem_two} we have $s \le b$ and $D \ge \sqrt{{s \over b} \cdot n \log n}$. In  \ref{subitem_two}, $s b \ge \sqrt{{s \over b} \cdot n \log n}$ implies $s b^3 \ge n\log n$, i.e.,
$b \le ((n \log n)/s)^{1/3}$.
In bound \ref{item:LargeD_one}, $s \le b \le ((n \log n)/s)^{1/3}$ implies $s b^3 \le s {{n \log n} \over s} = n \log n$. Hence the boundary of these two domains is $s b^3 = n \log n$. This (together with $b \ge s$) implies $b \ge (n \log n)^{1/4}$. 
In bound \ref{item:LargeD_one}, $s b = {{n \log n} \over {b^3}} \cdot b \ge D \ge \sqrt{{s \over b} \cdot n \log n}$. Since $s = {{n \log n} \over {b^3}}$, the right-hand side is equal to ${{n \log n} \over {b^2}}$, i.e., $D = {{n \log n} \over {b^2}}$.
Hence the intersection of these two domains is $D = {{n \log n} \over {b^2}}$. For this value of the diameter, the two bounds indeed agree, as
$T_0 = T_3 = {{n \log n} \over b}$.  

Similarly, the bounds  \ref{item:LargeD_one}  and \ref{subitem_three} also intersect when $D = {{n \log n} \over {b^2}}$ 
(i.e., $b = \sqrt{{n \log n} \over D}$) and $s = b$. Then 
$$\left({D \over b}\right)^{1/3} (n \log n)^{2/3} ~=~ \left({{D^{3/2}} \over {\sqrt{n \log n}}}\right)^{1/3} (n \log n)^{2/3} ~=~
\sqrt{D n \log n}~,$$ i.e., these two bounds agree on the boundary as well.

The bound \ref{item:LargeD_one} of Theorem \ref{thm:large_band_largeD} generalizes the single-source bound \ref{item:LargeD_three} from Theorem \ref{thm:single_src_large_band}.
The bound \ref{item:LargeD_three}  of  Theorem \ref{thm:large_band_largeD} generalizes 
the large-diameter multi-source bound of Theorem \ref{thm:few_srcs}. 

Next we compare the bounds of Theorem \ref{thm:large_band_largeD} with  the trivial bound of $O(\min \{n \lceil {s \over b} \rceil, D + {{ns+|E|} \over b}\})$. First, observe that this trivial bound can be improved only when $D \le {{|E| + ns} \over b}$, as these problems require $\Omega(D)$ time.  

The bound \ref{item:LargeD_one} improves the trivial bound (moreover, it is $o(ns/b)$) for $D = o\left({n \over {\log^2 n}} \cdot {{s^3}  \over {b^2}}\right)$.  
It is $o(n \lceil s/b\rceil) = o(n)$ (because $s \le b$ in bound \ref{item:LargeD_one}) for $D = o\left({{n \cdot b} \over {\log^2 n}}\right)$.


The bound \ref{item:LargeD_three} improves the trivial one for $D = o\left({{n \cdot s^2} \over {b \cdot {\log^2 n} }}\right)$.
Finally, the bound \ref{item:LargeDn} improves the trivial one for $D = o\left({n \over {\log n}} \cdot {{s^2} \over {b^2}}\right)$.
}

\newpage
\bibliographystyle{alpha}
\bibliography{spt}

\clearpage
\pagenumbering{roman}
\appendix      
\centerline{\LARGE\bf Appendix}

\section{Some Proofs}
\label{app:pf}

\inline Proof of Theorem \ref{thm:hopset}:
For a pair $u,v \in V$ of vertices, let $\pi(u,v) = (u = x_0,x_1,\ldots,x_\ell=v)$ be the shortest $u-v$ path with the smallest number of hops in $G \cup G^{(k)} = (V,E \cup H,\homega)$, $\homega(e) = \omega^{(k)}(e)$ for $e \in H$, and $\homega(e) = \omega(e)$ for $e \in E \setminus H$. 
(We assume that $\pi(u,v)$ has finite length, i.e., for every index $i$, $0 \le i \le \ell-1$, $\homega(x_i,x_{i+1}) < \infty$.)

We argue that for any index $i$, $0 \le i \le \ell/4-1$, we have $S[k](x_{4i}) \cap S[k](x_{4i+4}) = \emptyset$.

Observe that $x_{4i+2} \nin S[k](x_{4i})$, because otherwise we could obtain a shortest $u-v$ path in $G \cup G^{(k)}$ with fewer edges than in $\pi(u,v)$ by replacing the two edges $(x_{4i},x_{4i+1}),(x_{4i+1},x_{4i+2})$ in $\pi(u,v)$ by $(x_{4i},x_{4i+2})$.
Analogously, $x_{4i+2} \nin S[k](x_{4i+4})$  too.

Since $d_G(x_{4i},x_{4i+2}),d_G(x_{4i+2},x_{4i+4}) < \infty$, it follows that $|S[k](x_{4i})| = |S[k](x_{4i+4})| = k$.

Hence for any $y \in S[k](x_{4i})$,
\begin{equation}
\label{eq:xi}
d_G(x_{4i},y) \le d_G(x_{4i},x_{4i+2})~.
\end{equation}
Also, for any $y   \in S[k](x_{4i+4})$,
\begin{equation}
\label{eq:xifour}
d_G(x_{4i+4},y) \le d_G(x_{4i+4},x_{4i+2})~.
\end{equation}
So, if $S[k](x_{4i}) \cap S[k](x_{4i+4}) \neq \emset$, then there exists a vertex $y$ that satisfies both (\ref{eq:xi}) and (\ref{eq:xifour}).
But then one could get a shortest $u-v$ path in $G \cup G^{(k)}$ with fewer edges than in $\pi(u,v)$ by replacing the four edges of the subpath
$(x_{4i},\ldots,x_{4i+4})$ of $\pi(u,v)$ by the two edges $(x_{4i},y),(y,x_{4i+4})$, contradiction.

Consider the disjoint union $\bigcup_{i=0}^{\ell'} S[k](x_{4i})$, for the maximum $\ell'$ such that $4\ell' \le \ell$. It contains $(\ell'+1)k \le n$ vertices. Thus
$4n/k \ge 4(\ell'+1) > \ell$, as required.
\QED

\inline proof of Lemma \ref{lm:hop_constr}:
\Pfhopconstr

\inline proof of Lemma \ref{lm:nest}:
\Pfnest

\inline proof of Lemma \ref{lm:sqrt}:
\Pfsqrt

\inline proof of Lemma \ref{lm:Stagtag}:
\Pfstag

\section{Upcast}
\label{sec:upcast}

Consider a problem in which we are given a tree $\tau$ of hop-diameter $D$, rooted at a vertex $\rt$, and some $m$ distinct messages distributed in its vertices. (A single vertex may hold more than one message.) We want to {\em upcast} all these messages to the root $\rt$ of $\tau$ in the $\CONGEST(b \log n)$ model. When $b = 1$, this problem was analyzed in \cite{Pel00:ln}, Chapter 4, where it was shown that this task can be performed in $O(D + m)$ time. Here we extend this result to a general $b$, and show that in general, $O(D + m/b)$ time is sufficient. The argument follows closely that of \cite{Pel00:ln}; it is provided here for the sake of completeness.

The algorithm is a trivial one: whenever a vertex $v$ has some $q$ messages that it still did not send to its parent $p(v) = w$, it sends some arbitrary $\min\{q,b\}$ of them to $w$ (assuming that the edge $(v,w)$ is available for sending messages at this point).

For a vertex $v$, let $M(v)$ denote the set of messages initially stored in the subtree $\tau_v$ rooted at $v$. Let $m(v) = |M(v)|$.
 Let $\hL(v)$ denote the depth of $\tau_v$,  i.e., the maximum hop-distance between a leave $z$ in $\tau_v$ and its root $v$.

We start from the following simple lemma.

\begin{lemma}
\label{lm:aux}
Given a vertex $v$ and two positive integers $t,h$, {\em suppose} that for all $i$, $1 \le i \le h$, after $t + i$ rounds, the vertex $v$ has at least $\min\{ i \cdot b, m(v)\}$ messages. Then after $t+h+1$ rounds, the parent $w= p(v)$ of $v$ has $\min\{h \cdot b,m(v)\}$ messages received from $v$.
\end{lemma}
The proof of this lemma is  by a straightforward induction on $h$. We omit it.

\begin{lemma}
\label{lm:upcast}
For any vertex $w$, and for all $i$, $1 \le i \le \lceil {{m(w)} \over b} \rceil$, after $\hL(w) + i - 1$ round, at least $\min\{m(w), b \cdot i\}$ messages are at $w$.
\end{lemma}
\proof
The proof is by induction on $\hL(w)$. The base case is when $w$ is a leaf; it is immediate.

\inline Step: Suppose that the lemma holds for every child $v_j$ of $w$. Denote $\ell = \hL(w)$, $\ell_j = \hL(v_j)$, and $m_j = m(v_j)$, for all $1 \le j \le p$, where $p$ is the number of children of $w$. 
Define $\gamma_j = \min\{i, \lceil {{m_j} \over b}\rceil\} \le \lceil {{m_j} \over b} \rceil$. 

By the induction hypothesis for $v_j$, for every index $i'$, $1 \le i' \le \lceil {{m_j} \over b} \rceil$, after $\ell_j + i' - 1 = (\ell_j - 1) + i'$ rounds, $v_j$ has $\min\{i' \cdot b,m_j\}$ items.

Use Lemma \ref{lm:aux} with $t = \ell_j - 1$, $h = \gamma_j$. Since $\gamma_j \le \lceil {{m_j} \over b} \rceil$, the assumption of Lemma \ref{lm:aux} holds with these $t$ and $h$. Hence after $(\ell_j -1) + (\gamma_j +1) = \ell_j + \gamma_j$ rounds, the parent $w$ of $v_j$ has received at least $\min \{\gamma_j \cdot b,m(v_j)\} = m_j$ messages from $v_j$.

If there exists a child $v_j$ of $w$ with $\gamma_j \cdot b \le m_j$, then $w$ has received just from this child $v_j$ at least $\gamma_j \cdot b \ge i \cdot b \ge \min\{m(w), i\cdot b\}$ messages by time 
$$\ell_j + \gamma_j ~\le~ \hL(w) - 1 + \gamma_j ~\le~ \hL(w) - 1 + \lceil {{m_j} \over b} \rceil ~\le~ \hL(w) - 1 + \lceil {{m(w)} \over b} \rceil~,$$ as required.

Otherwise, every child $v_j$ of $w$ satisfies $ \gamma_j \cdot b > m_j$.
Hence $\gamma_j = \min\{i, \lceil {{m_j} \over b} \rceil\} > {{m_j} \over b}$.  It follows that $ i \ge \lceil {{m_j} \over b} \rceil = \min\{i, \lceil {{m_j} \over b} \rceil\} = \gamma_j$. 

Then, after $\max\{\ell_j + \gamma_j\} \le \max\{\ell_j\} + \max\{ \gamma_j\} ~\le~ \ell-1 + i$ rounds, the parent $w$ has received from each $v_j$ at least $\min\{m_j,b \cdot \lceil {{m_j} \over b} \rceil\} = m_j$ messages. Hence, by this time, $w$ has received all $m(w)$ messages from all its children. (Except for the messages that were originally stored at $w$, and they are kept being stored there.) 
\QED

Using the lemma with $w = r$ and $i = \lceil m(r)/b \rceil$, we conclude that all messages are collected at the root within $D + \lceil m(rt)/b \rceil$ rounds. They can, of course, be also disseminated to all vertices of the graph via an analogous pipelined broadcast (see  \cite{Pel00:ln}, Chapter 4) within the same time.


\APPlarge_band

\end{document}

%% file: latexmac.tex
\def\denseformat{
\setlength{\textheight}{9in}
\setlength{\textwidth}{6.9in}
\setlength{\evensidemargin}{-0.2in}
\setlength{\oddsidemargin}{-0.2in}
\setlength{\headsep}{10pt}
\setlength{\topmargin}{-0.3in}
\setlength{\columnsep}{0.375in}
\setlength{\itemsep}{0pt}
}

\newtheorem{theorem}{Theorem}[section]

\newtheorem{lemma}[theorem]{Lemma}

\newtheorem{corollary}[theorem]{Corollary}

\newtheorem{comment}[theorem]{Comment}

\def\boldhead#1:{\par\vskip 7pt\noindent{\bf #1:}\hskip 10pt}
\def\ithead#1:{\par\vskip 7pt\noindent{\it #1:}\hskip 10pt}

\def\inline#1:{\par\vskip 7pt\noindent{\bf #1:}\hskip 10pt}
\def\midinline#1:{\par\noindent{\bf #1:}\hskip 10pt}
\def\dnsinline#1:{\par\vskip -7pt\noindent{\bf #1:}\hskip 10pt}
\def\ddnsinline#1:{\newline{\bf #1:}\hskip 10pt}
\def\largeinline#1:{\par\vskip 7pt\noindent{\large\bf #1:}\hskip 10pt}
\long\def\comment #1\commentend{}
\long\def\commhide #1\commhideend{}
\long\def\commfull #1\commend{#1}
\long\def\commabs #1\commenda{}
\long\def\commtim #1\commendt{#1}
\long\def\commb #1\commbend{}
\long\def\commedit #1\commeditend{} 

\long\def\commB #1\commBend{}       

\long\def\commex #1\commexend{}     

\long\def\commsiena #1\commsienaend{}  
                                         
\long\def\commBI #1\commBIend{}  
                                         
\long\def\CProof #1\CQED{}

\def\blackslug{\hbox{\hskip 1pt \vrule width 4pt height 8pt
    depth 1.5pt \hskip 1pt}}
\def\QED{\quad\blackslug\lower 8.5pt\null\par}

\def\Proof{\par\noindent{\bf Proof:~}}

\def\proof{\Proof}

\long\def\PPP#1{\noindent{\bf Proof:}{ #1}{\quad\blackslug\lower 8.5pt\null}}

\long\def\denspar #1\densend
{#1}

\newif\ifnotesw\noteswtrue
   {\ifnotesw\marginpar[\hfill\(\top\)]{\(\top\)}\fi}%
   {\ifnotesw\marginpar[\hfill\(\bot\)]{\(\bot\)}\fi}

\newcommand{\mnote}[1]%
    {\ifnotesw\marginpar%
        [{\scriptsize\it\begin{minipage}[t]{\marginparwidth}
        \raggedleft#1%
                        \end{minipage}}]%
        {\scriptsize\it\begin{minipage}[t]{\marginparwidth}
        \raggedright#1%
                        \end{minipage}}%
    \fi}

\def\cA{{\cal A}}
\def\cB{{\cal B}}

\def\cN{{\cal N}}

\def\cP{{\cal P}}

\def\cS{{\cal S}}

\def\hL{{\hat L}}

\def\hi{{\hat i}}

\def\tC{{\tilde C}}

\def\tO{{\tilde O}}

\def\MathF{\hbox{\rm I\kern-2pt F}}
\def\MathP{\hbox{\rm I\kern-2pt P}}
\def\MathR{\hbox{\rm I\kern-2pt R}}
\def\MathZ{\hbox{\sf Z\kern-4pt Z}}
\def\MathN{\hbox{\rm I\kern-2pt I\kern-3.1pt N}}
\def\MathC{\hbox{\rm \kern0.7pt\raise0.8pt\hbox{\footnotesize I}
\kern-4.2pt C}}
\def\MathQ{\hbox{\rm I\kern-6pt Q}}

\def\MathE{\hbox{{\rm I}\hskip -2pt {\rm E}}} 

\newsavebox{\ttop}\newsavebox{\bbot}

\def\eps{\epsilon}

\def\polylog{\mbox{polylog}}

\def\setmns{\setminus}

\def\nin{{~\not \in~}}
\def\emset{\emptyset}

\newcommand{\Prob}{\MathP}
\newcommand{\Expect}{\MathE}

\def\etal{\emph{et~al.}}
